\documentclass[english]{article}
\usepackage{lmodern}

\usepackage[T1]{fontenc}
\usepackage[a4paper]{geometry}
\usepackage{color}
\usepackage{amsmath}
\usepackage{amsthm}
\usepackage{amssymb}
\usepackage{mathtools}
\usepackage{graphicx}
\usepackage{setspace}
\usepackage{caption}
\usepackage{natbib}
\usepackage[unicode=true,bookmarks=true,bookmarksnumbered=false,bookmarksopen=false,breaklinks=false,pdfborder={0 0 0},pdfborderstyle={},backref=page,colorlinks=true]{hyperref}

\def\equationautorefname~#1\null{(#1)\null}

\usepackage{authblk}
\hypersetup{
    pdftitle={Unbiased path sampling},
    pdfauthor={Rischard Jacob Pillai},
    allcolors=RoyalBlue
}

\author[1]{Maxime Rischard\thanks{\href{mailto:mrischard@g.harvard.edu}{mrischard@g.harvard.edu}}}
\author[1]{Pierre E. Jacob}
\author[1]{Natesh Pillai}
\affil[1]{Department of Statistics, Harvard University}

\usepackage{float}
\usepackage{subcaption}
\usepackage[dvipsnames,svgnames,x11names,hyperref]{xcolor}
\usepackage{nicefrac}
\usepackage{algorithm}

\floatstyle{ruled}
\newfloat{algorithm}{tbp}{loa}
\providecommand{\algorithmname}{Algorithm}
\floatname{algorithm}{\protect\algorithmname}

\providecommand{\assumptionname}{Assumption}

\newcommand{\setX}{\mathbb{X}}
\newcommand{\reals}{\mathbb{R}}
\DeclarePairedDelimiter{\parenthesis}{\lparen}{\rparen}
\DeclarePairedDelimiter{\squarebracket}{\lbrack}{\rbrack}
\DeclarePairedDelimiter{\curlybracket}{\lbrace}{\rbrace}
\DeclarePairedDelimiter{\absolutevalue}{\lvert}{\rvert}
\DeclarePairedDelimiter{\doublebar}{\lVert}{\rVert}
\newcommand{\del}[1]{\parenthesis*{#1}}
\newcommand{\sbr}[1]{\squarebracket*{#1}}
\newcommand{\cbr}[1]{\curlybracket*{#1}}
\newcommand{\abs}[1]{\absolutevalue*{#1}}
\newcommand{\norm}[1]{\doublebar*{#1}}
\newcommand*{\diffdchar}{d}
\newcommand*{\dif}[1]{\mathop{\diffdchar #1}}

\newcommand{\EMC}{\E_{\mathrm{MC}}}

\newcommand{\Esub}{\E_{\pathvar}}
\newcommand{\lambdadistr}{\lambda \sim \prob(\lambda)}
\newcommand{\Esup}{\E_{\lambdadistr}}
\newcommand{\probIS}{q}

\newcommand{\pathvar}{\lambda}
\newcommand{\paramvar}{x}
\newcommand{\paramrv}{X}
\newcommand{\covariate}{d}
\newcommand{\Covariate}{D}
\newcommand{\EU}{E}
\newcommand{\EUhat}{\hat{E}}

\newcommand{\pathgrid}[1]{\pathvar^{[#1]}}

\newcommand{\logratio}{r_{01}}
\newcommand{\hatratio}{\hat{r}_{01}}
\DeclareMathOperator{\prob}{\mathit{p}}
\DeclareMathOperator{\E}{\mathbb{E}}
\DeclareMathOperator{\V}{\mathbb{V}}

\makeatletter
\newcommand*{\iid}{%
    \@ifnextchar{.}%
        {i.\,i.\,d}%
        {i.\,i.\,d.\@}%
}
\makeatother

\newcommand{\pnorm}{\pi}
\newcommand{\unnorm}{\tilde\pi}
\newcommand{\barunnorm}{\bar\pi}

\DeclareMathOperator{\normal}{\mathcal{N}}
\DeclareMathOperator{\expo}{Exp}

\DeclareMathOperator{\invgamma}{\mathcal{IG}}
\DeclareMathOperator{\pg}{pg}

\newcommand{\eye}{\mathbf{I}}
\newcommand{\trans}{^{\intercal}}

\DeclareMathOperator{\expit}{expit}
\newcommand{\lagrange}{\xi}

\title{Unbiased estimation of log normalizing constants with applications to Bayesian cross-validation}

\begin{document}

\maketitle

\begin{abstract}
    Posterior distributions often feature intractable normalizing constants,
    called marginal likelihoods or evidence,
    that are useful for model comparison via Bayes factors.
    This has motivated a number of methods for estimating ratios of normalizing constants
    in statistics. In computational physics the logarithm of these ratios correspond to free energy differences.
    Combining unbiased Markov chain Monte Carlo estimators with path sampling, also called thermodynamic integration,
    we propose new unbiased estimators of the logarithm of ratios of normalizing constants.
    As a by-product, we propose unbiased estimators of the Bayesian cross-validation criterion.
    The proposed estimators are consistent, asymptotically Normal and can easily benefit from parallel processing
    devices.
    Various examples are considered for illustration.
\end{abstract}

\section{Setting \label{sec:setting}}

Monte Carlo methods address the approximation of intractable integrals of the form
$\pnorm(h) = \int h(\paramvar) \pnorm(\dif\paramvar)$, where $\pnorm$ is a probability distribution on a space $\setX$, for instance a subset of $\reals^d$, and $h$ a test function of interest.
In Bayesian inference $\pnorm$ is the posterior distribution  that combines the prior density $\paramvar \mapsto \prob(\paramvar )$
and the likelihood $\paramvar \mapsto \prob(y \mid \paramvar )$, where the data $y$ are assumed fixed, through the relation $\pnorm(\paramvar ) = \prob(\paramvar ) \prob(y \mid \paramvar ) / \prob(y)$ where $\prob(y) = \int_\setX \prob(y \mid \paramvar )\prob(\paramvar )\dif\paramvar$.
Using various Markov chain Monte Carlo methods \citep{robert:casella:2004,stoltz2010free,brooks2011handbook}
one can approximate $\pnorm(h)$ without having access to the normalizing constant $\prob(y)$, which is often intractable.
However, the interest is sometimes in the normalizing constant $\prob(y)$ itself \citep{chen1997monte}, also called the marginal likelihood or evidence,
as it can be used for model comparison \citep[e.g.][]{jeffrey1939,bernardo2001bayesian,dawid2011}.
Below we denote the normalizing constant of $\pnorm$ by $Z$, and
the unnormalized density by $\unnorm$, so that $\pnorm(\paramvar )=\unnorm(\paramvar )/Z$.

In this article we propose a new estimator of $Z$, which combines unbiased
Markov chain Monte Carlo \citep{jacob2017unbiased} with the path sampling
identity (\citet{gelman1998simulating}; see also Chapter~5 of \citet{chen2000monte}), 
also known as thermodynamic integration
\citep{kirkwood1935statistical,neal2005estimating,calderhead2009estimating}.
The specificity of
the proposed estimator is its unbiasedness for the logarithm of $Z$, i.e. the
expectation of the proposed estimator is exactly $\log Z$.
Existing estimators based on Markov chain Monte Carlo \citep{chen1997monte} are only asymptotically unbiased,
while existing estimators based on annealed importance samplers
\citep{neal:2001} and sequential Monte Carlo samplers \citep{DelDouJas:2006}
are unbiased for $Z$ and not for $\log Z$.

Leveraging unbiasedness for $\log Z$, we consider a
Bayesian cross-validation (CV) criterion based on the logarithmic scoring rule
\citep[e.g][]{alqallaf2001cross,bornn2010efficient,vehtari2017practical}.
In
cross-validation, one randomly splits the available data into training and
validation, then the posterior distribution given the training data is
numerically approximated, and finally the predictive performance on the validation data is
assessed e.g. with the logarithmic scoring rule \citep{parry2012}.
We
propose an estimator that is directly unbiased for these Bayesian cross-validation
objectives, which can be averaged over independent copies to obtain consistent estimators 
and asymptotically exact confidence intervals from the central limit theorem for \iid{} variables.

The rest of the document is structured as follows.
\autoref{sec:methodology}
introduces the proposed estimators, and their tuning parameters are discussed.
Numerical experiments in simple examples can be found in \autoref{sec:numerics}.
\autoref{sec:discussion} discusses our findings and future directions. The code
to reproduce the experiments of the article is available at
\url{https://github.com/pierrejacob/unbiasedpathsampling}.

\section{Proposed estimators \label{sec:methodology}}

We propose an unbiased estimator of $\log Z$ in \autoref{sec:unbiasedpathsampling},
and obtain an unbiased estimator of a Bayesian cross-validation criterion in \autoref{sec:bayesiancrossvalidation}.
Our implementation relies on the unbiased MCMC estimators of \citet{jacob2017unbiased},
which are briefly reviewed in \autoref{sec:unbiasedMCMC},
while \autoref{sec:tuning} discusses tuning choices.

\subsection{Unbiased path sampling \label{sec:unbiasedpathsampling}}

We first recall thermodynamic integration, or path sampling,
for the approximation of normalizing constants
\citep{chen1997monte,gelman1998simulating,calderhead2009estimating,cameron2014recursive},
see also \citet{stoltz2010free} for a thorough overview of related methods.
We introduce a ``path'' of distributions:
$\pnorm_\pathvar(\paramvar)=\exp(-U_{\pathvar}(\paramvar ))/Z_{\pathvar}$, with $\pathvar\in[0,1]$, and
$Z_\pathvar = \int \exp(-U_{\pathvar}(\paramvar ))\dif\paramvar$.
We also write $\unnorm_\pathvar(\paramvar ) = \exp(-U_\pathvar(\paramvar ))$.
The path is such that the object of
interest is $\logratio = \log(Z_1/Z_0)$.
For instance, it could represent the difference in the logarithm of
the marginal likelihood (or evidence) between two models in a Bayesian setting.
In settings where
$U_0$ and $U_1$ are given, a common example of path is the ``geometric'' path defined as
$U_{\pathvar}:x\;\mapsto (1-\pathvar)U_{0}(\paramvar ) + \pathvar U_{1}(\paramvar )$ for
all $\pathvar\in[0,1]$.
The geometric path is not optimal in any way
but it can be practical; we will also discuss other choices in the experiments.

The thermodynamic integration or path sampling identity relies on the following
interchange between differentiation and integration \citep{kirkwood1935statistical},
\begin{equation}
    \nabla_\pathvar \log Z_{\pathvar}
    =
    \frac{
        \nabla_\pathvar Z_{\pathvar}
    }{
        Z_{\pathvar}
    }
    =
    \frac{
        \nabla_\pathvar
        \sbr{
            \int\unnorm_{\pathvar}(\paramvar )\dif\paramvar
        }
    }{
        Z_{\pathvar}
    }
    =
    \int
        \nabla_\pathvar\sbr{
            \log\unnorm_{\pathvar}(\paramvar)
        }
        \:
        \pnorm_\pathvar(\paramvar)
        \dif\paramvar
    \,,
\end{equation}
where $\nabla_\pathvar$ denotes derivative with respect to $\pathvar$.
The formula holds under regularity conditions such as: $\pathvar \mapsto \nabla_\pathvar \unnorm_\pathvar(\paramvar )$ is continuous
for all $\paramvar $, and there exists an integrable function $\paramvar \mapsto \barunnorm(\paramvar )$ such that $\abs{\nabla_\pathvar \unnorm_\pathvar(\paramvar )} \leq \barunnorm(\paramvar )$ for all $\paramvar $ and for all $\pathvar$.
Denoting by $\Esub$ expectations with respect to $\pnorm_\pathvar$,
integrating the above expression with respect to $\pathvar$
yields
\begin{equation}
    \logratio = \log Z_{1}-\log Z_{0}= - \int_{0}^{1}\Esub[ \nabla_\pathvar U_\pathvar(\paramrv) ] d\pathvar.
\end{equation}
By introducing an arbitrary density $\pathvar\mapsto \probIS(\pathvar)$, strictly
positive on $(0,1)$, we obtain the path sampling identity:
\begin{equation}
    \logratio = - \int_{0}^{1}\frac{\Esub[\nabla_\pathvar U_{\pathvar}(\paramrv)]}{\probIS(\pathvar)}\probIS(\pathvar)d\pathvar.\label{eq:pathsamplingidentity}
\end{equation}
This is useful if we can approximate integrals with respect to $\pathvar$
by Monte Carlo or numerical integration, and if we can approximate the
inside expectation $\Esub[\nabla_\pathvar U_\pathvar(\paramrv)]$ by Markov chain
Monte Carlo (MCMC, \citet{robert:casella:2004}), for instance.

For instance, we might discretize $\pathvar$ on $[0,1]$ by introducing a grid of points $\pathgrid{1},\dotsc,\pathgrid{L}$.
Then for each $l \in \{1, \dotsc, L\}$ and $\pathvar = \pathgrid{l}$, we could approximate each $\E_{\pathvar}[\nabla_\pathvar U_\pathvar(\paramrv)]$ with an MCMC estimator based on $T_l$ iterations.
We could finally aggregate these estimators to obtain a consistent estimator for $\logratio=\log(Z_1/Z_0)$,
as $L \to \infty$ and as $T_l\to\infty$ for all $l\in \{1,\dotsc,L\}$ \citep{gelman1998simulating}.
Instead, if we directly define an MCMC algorithm targeting the distribution $\probIS(\dif\pathvar)\pnorm_\pathvar(\dif\paramvar)$
on the joint space $[0,1]\times \setX$, then we can obtain an estimator of $\logratio$ that would be valid
in a single asymptotic regime, as the number of iterations goes to infinity.

Here we denote by $\EU(\pathvar) = -\E_\pathvar[\nabla_\pathvar U_\pathvar(\paramrv)]$ the inner expectation
in \autoref{eq:pathsamplingidentity}, and we introduce $\EUhat(\pathvar)$,
an unbiased estimator of $\EU(\pathvar)$ that we can generate for any $\pathvar\in[0,1]$;
we defer the construction of such estimators to \autoref{sec:unbiasedMCMC}.
We can then define an estimator of $\logratio=\log (Z_1/Z_0)$ with the following procedure.
\begin{enumerate}
    \item Draw $\pathvar\sim \probIS(\dif\pathvar)$, a distribution supported on $[0,1]$.
    \item Given $\pathvar$, generate a variable $\EUhat(\pathvar)$ with expectation $\EU(\pathvar) = -\E_\pathvar[\nabla_\pathvar U_\pathvar(\paramrv)]$.
    \item Return $\hatratio = \EUhat(\pathvar) / \probIS(\pathvar)$.
\end{enumerate}
The random variable $\hatratio$ has expectation $\logratio$ by the law of iterated expectations,
and we refer to it as an unbiased path sampling estimator (UPS).
Note that sequential Monte Carlo samplers and related methods \citep{DelDouJas:2006}
would provide unbiased estimators of $Z$ and not of $\log Z$.
Thus these estimators will not be unbiased for $\logratio$.
We will now see that the lack of bias on the logarithmic
scale can be exploited to propose new estimators of Bayesian cross-validation criteria.

\subsection{Unbiased Bayesian cross-validation \label{sec:bayesiancrossvalidation}}

A number of articles discuss the computational difficulties associated with Bayesian cross-validation, e.g.
\citet{alqallaf2001cross,bhattacharya2007importance,bornn2010efficient,lamnisos2012cross,mcvinish2013recentered,vehtari2017practical}.
We first define the object of interest, before presenting our estimator.
Let $\paramvar$ denote an unknown parameter with prior density $\prob(\paramvar)$,
and let $y_{1:n} = \{y_1,\dotsc,y_n\}$ denote the data composed of $n$ units.
The likelihood function is denoted by $\paramvar\mapsto \prob(y_{1:n} \mid \paramvar)$.
Cross-validation consists in randomly splitting $y_{1:n}$ into $T$ and $V$, where
$T$ stands for training and $V$ for validation.
The sets $T,V$ form a partition of $y_{1:n}$, $T\cap V = \emptyset$ and $T\cup V = y_{1:n}$.
Denote by $n_T$ and $n_V$ the numbers of elements in $T$ and $V$; for instance, if $n_T = n-1$,
the procedure is termed ``leave-one-out'' cross-validation.
Given a split of the data $y_{1:n} = (T,V)$, we introduce
a measure of accuracy in predicting $V$ using the training data $T$.
A typical choice is the logarithmic score $-\log \prob(V \mid T)$ \citep[see][for a discussion on the choice of scoring rule]{parry2012} where
$\prob(V \mid T) = \int \prob(V \mid T, \paramvar)\prob(\paramvar \mid T)\dif\paramvar$
is the posterior predictive density given $T$ and evaluated on $V$.
Note that $\prob(V \mid T, \paramvar)$ simplifies to $\prob(V \mid \paramvar)$
if the data are modeled as conditionally independent given $\paramvar$.
The cross-validation objective, ``CV'' below, is defined as an average over all splits $(T,V)$ of size $(n_T,n_V)$,
\begin{equation}
    \mathrm{CV} = - \binom{n}{n_T}^{-1} \sum_{T,V \in \mathcal{S}} \log \prob(V \mid T),
    \label{eq:cvobjective}
\end{equation}
where $\mathcal{S}$ is the set of partitions of $\{1,\dotsc,n\}$ into $T,V$ of sizes
$n_T,n_V$.
To approximate this criterion, one can sample partitions $T,V$,
and approximate $\log \prob(V \mid T)$ with MCMC estimators.
For any fixed $n_T$, this procedure
would give consistent estimates of $\mathrm{CV}$ as the number of splits and the number of MCMC iterations go to infinity.

Given a split $T,V$, we can estimate $\log \prob(V \mid T)$ using the path sampling identity
and the unbiased estimators of the previous section.
Indeed, that quantity is a log-ratio of the
normalizing constants $\prob(T)$ and $\prob(T,V)$.
By introducing the path
\begin{equation}
\forall \: \pathvar \in [0,1] \quad \unnorm_\pathvar(\paramvar) = \prob(\paramvar) \prob(T \mid \paramvar) \prob(V \mid T,\paramvar)^\pathvar,
    \label{eq:cvsequence}
\end{equation}
we have $Z_0 = \int \unnorm_0(\paramvar) \dif\paramvar = \prob(T)$ and $Z_1 = \int \unnorm_1(\paramvar) \dif\paramvar =
\prob(T,V)$, thus $\log(Z_1/Z_0) = \log \prob(V \mid T)$.
Other paths can be used, as long as $Z_0 = \prob(T)$ and $Z_1 = \prob(T,V)$.
Assuming that we
can perform unbiased MCMC targeting $\pnorm_\pathvar$ for all $\pathvar$, and that we
can evaluate $\nabla_\pathvar \log \unnorm_\pathvar(\paramvar)$ for all $\paramvar,\pathvar$, then
we can obtain unbiased estimators of $\log \prob(V \mid T)$.

This motivates the following strategy: sample a split $T,V$ uniformly from $\mathcal{S}$,
and then obtain an unbiased estimator of $-\log \prob(V \mid T)$ given $T,V$.
The resulting estimator is
directly unbiased for CV in \autoref{eq:cvobjective}, by the law of iterated
expectations.
We summarize the procedure below.

\begin{enumerate}
    \item Sample index sets $T,V$ uniformly at random over $\mathcal{S}$, the set of partitions of $\{1,\dotsc,n\}$ into
        a set of size $n_T$ and a set of size $n_V = n-n_T$.
    \item Given $T,V$, introduce a path $(\unnorm_\pathvar)$ with $\pathvar\in[0,1]$, with constant $Z_\pathvar = \int \unnorm_\pathvar(\paramvar)\dif\paramvar$ such that
        $\logratio = \log(Z_1 / Z_0) = \log \prob(V \mid T)$.
Given the
        path, obtain an unbiased estimator  of $\log \prob(V \mid T)$, denoted by $\hatratio$.
    \item Return $-\hatratio$, an unbiased estimator of $\mathrm{CV}$ in \autoref{eq:cvobjective}.
\end{enumerate}

Note how the lack of bias on the logarithmic scale is important for the above procedure to produce an unbiased estimator of CV.
We could also extend the above procedure to allow for non-uniform sampling of the partitions from $\mathcal{S}$.

\subsection{Reminders on unbiased MCMC \label{sec:unbiasedMCMC}}

The UPS algorithm of \autoref{sec:unbiasedpathsampling} presupposes the ability to unbiasedly estimate expectations of the form
$\pnorm(h):=\int h(\paramvar)\pnorm(\dif\paramvar)$,
where $\pnorm$ is a target distribution, and $h$ is a test function.
In this paper, we use unbiased estimators recently proposed in \citet{jacob2017unbiased},
themselves building on those in \citet{glynn2014exact},
though other unbiased estimators could be substituted.
We thus briefly recall the estimators proposed in \citet{jacob2017unbiased},
and the associated tuning parameters.
Introduce a Markov kernel $P$, i.e. $P(\paramvar,\cdot)$ is a distribution on $\setX$ for all $\paramvar\in \setX$,
and for any measurable set $A$, the function $\paramvar \mapsto P(\paramvar,A)$ is measurable, and assume that $P$ is $\pnorm$-invariant.
Next, introduce a ``coupled'' Markov kernel $\bar{P}$ on the joint space $\setX\times \setX$,
such that for all $\paramvar,\tilde{\paramvar},A,B$,
$\bar{P}((\paramvar,\tilde{\paramvar}),(A,\setX)) = P(\paramvar,A)$ and
$\bar{P}((\paramvar,\tilde{\paramvar}),(\setX, B)) = P(\tilde{\paramvar},B)$,
i.e. $\bar{P}$ couples $P$ with itself.
Furthermore we will construct $\bar{P}$ such that,
at least for certain pairs $(\paramvar,\tilde{\paramvar})$, the distribution
$\bar{P}((\paramvar,\tilde{\paramvar}),\cdot)$ puts some non-zero mass on the diagonal
$\{(\paramvar',\tilde{\paramvar}')\in\setX\times \setX: \paramvar'=\tilde{\paramvar}'\}$.

With these elements, introduce two Markov chains $(X_{n})_{n\ge0}$
and $(\tilde{X}_{n})_{n\ge0}$ as follows.
First, $X_0$ and $\tilde{X}_0$ are drawn from an initial distribution (for simplicity, independently).
Then $X_{1}$ is sampled from $P(X_0,\cdot)$.
At step $n\geq1$, the pair $(X_{n+1},\tilde{X}_{n})$ is sampled
from the coupled kernel $\bar{P}((X_{n},\tilde{X}_{n-1}),\cdot)$.
The construction must be such that, for all $n\geq 0$, $X_n$ has the same distribution
as $\tilde{X}_n$, and such that there exists a random variable $\tau$,
referred to as the ``meeting time,''
such that for all $n\geq \tau$, $X_n = \tilde{X}_{n-1}$, almost surely.
We then introduce two integers, $k\geq 0$ and $m\geq k$, which will be tuning parameters,
and define the estimator
\begin{equation}
    H_{k:m}
    =
    \frac{1}{m-k+1}
    \sum_{n=k}^{m}
        h(X_{n})
    + \sum_{n=k+1}^{\tau-1}
        \min\del{1, \frac{n-k}{m-k+1} }
        \del{
            h(X_{n})
            -h(\tilde{X}_{n-1})
        }
    \,.
    \label{eq:bc}
\end{equation}
In the above expression, the convention is that the sum $\sum_{n=k+1}^{\tau-1}$ is equal to zero in the event $\tau-1<k+1$.
The estimator $H_{k:m}$ is a standard Markov chain
average $(m-k+1)^{-1}\sum_{n=k}^{m}h(X_{n})$ based on $m$ iterations and a burn-in of $k-1$ steps,
plus another term that is precisely such that $\EMC[H_{k:m}] = \pnorm(h)$,
where $\EMC$ denotes expectation with respect to all random variables involved in the Monte Carlo algorithm; see \citet{jacob2017unbiased} for
more precise statements.

\subsection{Tuning choices \label{sec:tuning}}

A number of choices have to be made for the proposed estimators to be operational.
The first choice is that of a path of distributions.
There are generic choices such as the geometric path, and choices motivated by algorithmic considerations on a case-by-case basis.
We will discuss the choice of paths through examples, in \autoref{sec:numerics}.

Given a path of distributions $(\pnorm_\pathvar)$, algorithms approximating expectations $\E_\pathvar$ with respect to $\pnorm_\pathvar$
typically involve tuning parameters.
We describe the tuning of unbiased MCMC in \autoref{sec:tuning:unbiasedestimators}.
Then we discuss choices of distribution $\probIS(\dif\pathvar)$ in
\autoref{sec:tuning:q}.

\subsubsection{Tuning of unbiased MCMC\label{sec:tuning:unbiasedestimators}}

The unbiased MCMC estimators described in Section \autoref{sec:unbiasedMCMC}
require the specification of a Markov kernel $P$, a coupled kernel $\bar{P}$, and
an initial distribution for the chains. Specifying these objects is typically difficult,
but not specific to the setting of normalizing constant estimation.
Therefore we defer to the large literature on MCMC algorithms \citep{robert:casella:2004,brooks2011handbook},
as well as the relevant discussions in \citet{jacob2017unbiased} in the context of unbiased MCMC.
Ultimately we will care about the expected cost and the variance of the proposed unbiased estimators,
in order to maximize the efficiency of the proposed estimators, as discussed in the next section.

We thus discuss the expected cost and variance of unbiased MCMC estimators.
Since the meeting time $\tau$ is a random variable, the cost of generating $H_{k:m}$ in \autoref{eq:bc} is random.
Neglecting the cost of drawing from the initial distribution, the cost amounts to that of one
draw from the kernel $P$, $\tau-1$ draws from the kernel $\bar{P}$,
and then $(m-\tau)$ draws from $P$ if $\tau<m$.
Overall that leads to an expected cost of $C:=\EMC[\tau-1+\max(\tau,m)]$ units,
where each unit is the cost of drawing from $P$, and assuming that one sample from
$\bar{P}$ costs two units.
Note that the expected cost is approximately $m+\EMC[\tau]$ when $m$ is much larger than typical values of $\tau$.
The guidelines for the choice of $k$ and $m$ in \citet{jacob2017unbiased}
are to set $k$ such that the probability of $\{\tau > k\}$ is small, based on draws of $\tau$.
Then $m$ can be set to be a multiple of $k$, such as $2k$ or $5k$, so that the proportion of discarded iterations
remains small. In \citet{jacob2017unbiased}, under further conditions on the Markov kernels,
it is shown that the variance of unbiased MCMC estimators is equivalent to the variance of standard MCMC estimators
when $k$ and $m$ are large enough. Informally this confirms that the increased variance incurred by the removal of the bias
can be inconsequential if we choose $k$ and $m$ carefully.

For our purposes, the test function $h$ will be $\paramvar\mapsto -\nabla_\pathvar U_\pathvar(\paramvar)$
and the target distribution $\pnorm_\pathvar$, for different $\pathvar \in [0,1]$.
We will index the meeting time $\tau$,
the integers $k$ and $m$ and the expected cost $C$ by $\pathvar$.
The corresponding estimator is denoted $\EUhat(\pathvar)$
and has expectation $\EU(\pathvar) = \E_\pathvar[-\nabla_\pathvar U_\pathvar(X)]$.
We also introduce notation for the second moment of $\EUhat(\pathvar)$: let $m_2(\pathvar) = \EMC[\EUhat(\pathvar)^2]$
for all $\pathvar\in[0,1]$.
We will assume that the kernels $P_\pathvar$ and $\bar{P}_\pathvar$ corresponding to each target $\pnorm_\pathvar$
are such that $m_2(\pathvar) < \infty$ and $C_\pathvar < \infty$ for all $\pathvar \in [0,1]$; see \citet{jacob2017unbiased} and \citet{middleton2018unbiased}
for assumptions on the kernels under which the second moment $m_2(\pathvar)$ and the cost $C_\pathvar$ are guaranteed to be finite.

\subsubsection{Tuning of the distribution $\probIS(\dif\pathvar)$ \label{sec:tuning:q}}

Given unbiased MCMC estimators $\EUhat(\pathvar)$ of $\EU(\pathvar)$ for all $\pathvar$,
we move on to the choice of probability density function $\pathvar \mapsto \probIS(\pathvar)$.
Various choices lead to valid estimators, provided that the support of $\probIS(\dif\pathvar)$ is the entire interval $[0,1]$,
but we might want to maximize the efficiency of the estimator $\hatratio$
of $\logratio = \log(Z_1/Z_0)$.
We introduce the inefficiency as the product of expected cost,
$\int C_\pathvar \probIS(\pathvar)\dif\pathvar$ and variance $\V[\hatratio]$,
motivated by \citet{Glynn1991,glynn1992asymptotic}.
The efficiency is defined as the inverse of the inefficiency.
The variance $\V[\hatratio]$ is equal to $\EMC[\hatratio^2]-\logratio^2$,
and $\hatratio = \EUhat(\pathvar)/\probIS(\pathvar)$ with $\pathvar \sim \probIS(\dif\pathvar)$,
thus $\EMC[\hatratio^2] = \int (m_2(\pathvar)/\probIS(\pathvar)) \dif\pathvar$,
which leads to the following optimization program over functions $q$,
\begin{equation}
    \begin{split}
        &\min_{q} \cbr{
            \int C_\pathvar \probIS(\pathvar)\dif\pathvar
            \;\times\;
            \del{
                \int \frac{m_2(\pathvar)}{\probIS(\pathvar)} \dif\pathvar - \logratio^2
            }
        }
       \label{eq:optimization} \\
        &\text{such that} \quad \int \probIS(\pathvar)\dif\pathvar = 1, \quad \text{and} \quad \forall \: \pathvar \in [0,1] \quad \probIS(\pathvar) \geq 0 \,.
    \end{split}
\end{equation}
The above program is simpler if the cost $C_\pathvar$ is constant over $\pathvar$.
This can be enforced by an appropriate choice of parameters $m_\pathvar$, since $C_\pathvar \approx m_\pathvar + \EMC[\tau_\pathvar]$.
Therefore we will choose $m_\pathvar$ to make $C_\pathvar$ approximately constant, based on preliminary draws of $\tau_\pathvar$ on a grid of values of $\pathvar$.

If $C_\pathvar$ is constant over $\pathvar$, then the solution of the above minimization problem is given by $\pathvar \mapsto q^\star(\pathvar)\propto \sqrt{m_2(\pathvar)}$.
In \citet{gelman1998simulating} that solution is given, and then
the verification that this is indeed a solution is done via Cauchy-Schwarz.
Here we provide an informal derivation of the solution,
in the case where $C_\pathvar$ is constant over $\pathvar$.
We write the function to minimize as
$\int m_2(\pathvar)/\probIS(\pathvar) \dif\pathvar$,
and introduce the Lagrangian
\begin{equation}
    \int \frac{m_2(\pathvar)}{\probIS(\pathvar)} \dif\pathvar + \lagrange \del{\int \probIS(\pathvar) \dif\pathvar - 1}.
\end{equation}
We would like to differentiate with respect to $q$ and set the derivative to zero.
Introduce the directional derivative $\frac{\dif{}}{\dif\varepsilon} (\probIS(\pathvar)+\varepsilon v (\pathvar))$ where $v(\pathvar)$ is a function.
Replacing $\probIS(\pathvar)$ by $\probIS(\pathvar)+\varepsilon v(\pathvar)$ and differentiating with respect to $\varepsilon$ in the Lagrangian yields
\begin{equation}
    \begin{aligned}
        & \frac{\dif{}}{\dif\varepsilon} \cbr{\int \frac{m_2(\pathvar)}{\probIS(\pathvar)+\varepsilon v(\pathvar)} \dif\pathvar + \lagrange \del{\int (\probIS(\pathvar)+\varepsilon v(\pathvar)) \dif\pathvar - 1}}\\
        & = \int \frac{\dif{}}{\dif\varepsilon} \cbr{\frac{m_2(\pathvar)}{\probIS(\pathvar)+\varepsilon v(\pathvar)} + \lagrange(\probIS(\pathvar)+\varepsilon v(\pathvar)) } \dif\pathvar \quad \text{(interchange deriv.
    and integr.)} \\
        & = \int \cbr{\frac{-m_2(\pathvar)v(\pathvar)}{(\probIS(\pathvar)+\varepsilon v(\pathvar))^2} + \lagrange v(\pathvar) } \dif\pathvar.
    \end{aligned}
\end{equation}
Setting $\varepsilon$ to zero yields
$\int \{
   -m_2(\pathvar)/\probIS(\pathvar)^2 + \lagrange
    \} v(\pathvar) \dif\pathvar$,
and trying to set that expression to zero simultaneously for every choice of $v$, we obtain $-m_2(\pathvar)/(\probIS(\pathvar)^2) + \lagrange = 0$, i.e.
$\probIS(\pathvar)\propto \sqrt{m_2(\pathvar)}$.
This gives the candidate solution.

\subsubsection{Proposed tuning procedure\label{sec:proposedtuningprocedure}}

We now combine the above sections into practical guidelines for the proposed estimators.

\begin{enumerate}
    \item Set up path, $\unnorm_\pathvar$ for $\pathvar\in[0,1]$, such that $\int \unnorm_0(\paramvar)\dif\paramvar = Z_0$ and $\int \unnorm_1(\paramvar)\dif\paramvar = Z_1$,
        and such that the object of interest is $\logratio = \log(Z_1/Z_0)$.
    \item For $\pathvar$ in a grid of $L+1$ values $0= \pathgrid{0} \leq \ldots \leq \pathgrid{L}= 1$, 
        construct and tune an unbiased MCMC (initial distribution, Markov kernel $P$, and coupled kernel $\bar{P}$) targeting $\pnorm_\pathvar$,
        and draw independent samples of the associated meeting times $\tau_\pathvar$.
    \item Based on the distribution of meeting times at each $\pathvar$, choose $k_\pathvar$ and $m_\pathvar$,
        to complete the tuning of the unbiased estimator $\EUhat(\pathvar)$.
        The choice of $m_\pathvar$ can be made such that the expected cost
        $C_\pathvar = \EMC[\tau_\pathvar-1+\max(\tau_\pathvar,m_\pathvar)]$
        is approximately constant over $\pathvar$.
    \item Draw independent samples of $\EUhat(\pathvar)$ using the chosen $k_\pathvar$ and $m_\pathvar$,
        and estimate $m_2(\pathvar) = \EMC[\EUhat(\pathvar)^2]$ for $\pathvar$ in the grid $\pathgrid{0} \leq \ldots \leq \pathgrid{L}$.
    \item Use these estimates to define a distribution $\probIS(\dif\pathvar)$,
        such that $\probIS(\pathvar)$ is approximately proportional to $\sqrt{m_2(\pathvar)}$
        for all $\pathvar$ in $[0,1]$.
\end{enumerate}
We describe a concrete way of performing step 5, for completeness.
Given a grid of values $0 = \pathgrid{0} \leq \ldots \leq \pathgrid{L} =  1$
and associated estimates $(\hat{m}_2(\pathgrid{l}))^{1/2}$
of $(m_2(\pathgrid{l}))^{1/2}$ for $l \in \{0,\dotsc,L\}$ obtained in step 4,
we can define a distribution $\probIS(\dif\pathvar)$ that is piecewise uniform 
on the intervals $[\pathgrid{l}, \pathgrid{l+1}]$,
and such that
\begin{equation}
    \forall \: l \in \{0, \dotsc, L-1 \}
    \quad
    \int_{\pathgrid{l}}^{\pathgrid{l+1}}
    \probIS(\dif\pathvar)
    \propto
    (\pathgrid{l+1} - \pathgrid{l})
    \times
    \frac{
        \sqrt{\hat{m}_2(\pathgrid{l})}
        +
        \sqrt{\hat{m}_2(\pathgrid{l+1})}
    }{
        2
    }.
\end{equation}
Sampling from such a distribution can be done in order $\mathcal{O}(\log L)$ operations,
by first selecting an interval $[\pathgrid{l},\pathgrid{l+1}]$ with probability $\int_{\pathgrid{l}}^{\pathgrid{l+1}} \probIS(\dif\pathvar)$, and then sampling uniformly from that interval.

After the preliminary phase described in the five steps above, the generation of estimators $\hatratio$ can proceed as follows.
First, $\pathvar$ is drawn from $\probIS(\dif\pathvar)$ obtained in step 5 above.
We then find the nearest value $\pathgrid{l}$ in the grid,
with index $l \in \{0,\dotsc,L\}$.
We can look up tuning parameters corresponding to $\pathgrid{l}$ for the unbiased MCMC estimators,
stored during step 2 above, and the values of $k_{\pathgrid{l}}$ and $m_{\pathgrid{l}}$ stored during step 3 above.
Using these tuning values
we can generate an unbiased estimator $\EUhat(\pathvar)$ of $\EU(\pathvar) = -\E_\pathvar[\nabla_\pathvar U_\pathvar(\paramrv)]$.
The estimator $\hatratio = \EUhat(\pathvar)/\probIS(\pathvar)$ is finally returned.

\section{Numerical experiments \label{sec:numerics}}

The numerical experiments are structured as follows.
Section
\ref{sec:numerics:constants} contains toy examples of unbiased path sampling estimators.
\autoref{sec:pgg} considers logistic regressions with different choices of paths
and of unbiased MCMC estimators, and an example taken from \citet{epifani2008case,vehtari2017practical}.
\autoref{sec:linearregression} considers linear regressions with examples
taken from \citet{alqallaf2001cross,peruggia1997variability,vehtari2017practical}.
Throughout the experiments, 95\% confidence intervals for an estimand $\mu$ are obtained as $\bar{\mu} \pm 1.96 s / \sqrt{M}$, 
where $\bar{\mu}$ is the mean of $M$ independent unbiased estimators of $\mu$ and  $s$ is their sample standard deviation.
These confidence intervals are justified asymptotically as $M\to\infty$ by the central limit theorem for \iid{} random variables,
provided that the variance of the unbiased estimators is finite. On parallel machines and under budget constraints, 
valid confidence intervals can be constructed following \cite{Glynn1991}; see also related remarks in \citet{jacob2017unbiased}.

\subsection{Toy examples of normalizing constant estimation \label{sec:numerics:constants}}

\subsubsection{Normal example \label{sec:normalexample}}

We start with the example of Section 4.4 in
\citet{gelman1998simulating}.
Consider $\unnorm_{\pathvar}(\beta)=\exp(-(\beta- \pathvar
D)^{2}/2)$, with $D = 4$, which corresponds to a sequence of distributions
$\pnorm_\pathvar$ that interpolates between $\normal(0,1)$ and
$\normal(D,1)$.
The normalizing constants are $Z_{\pathvar}=\sqrt{2\pi}$ for
all $\pathvar$, so that $\logratio=0$.
Here we have
$\nabla_\pathvar \log \unnorm_\pathvar:x\mapsto D(\beta-\pathvar D)$.
To estimate
expectations $\E_\pathvar$, we consider a Metropolis--Hastings (MH)
algorithm, starting from an initial distribution $\normal(-1,2^{2})$, and
with Normal random walk proposals with variance $1$.
We couple this
algorithm by maximally coupling the proposal distributions \citep{jacob2017unbiased}.

\begin{figure}
    \centering
    \begin{subfigure}[t]{0.3\textwidth}
     \includegraphics[width=\textwidth]{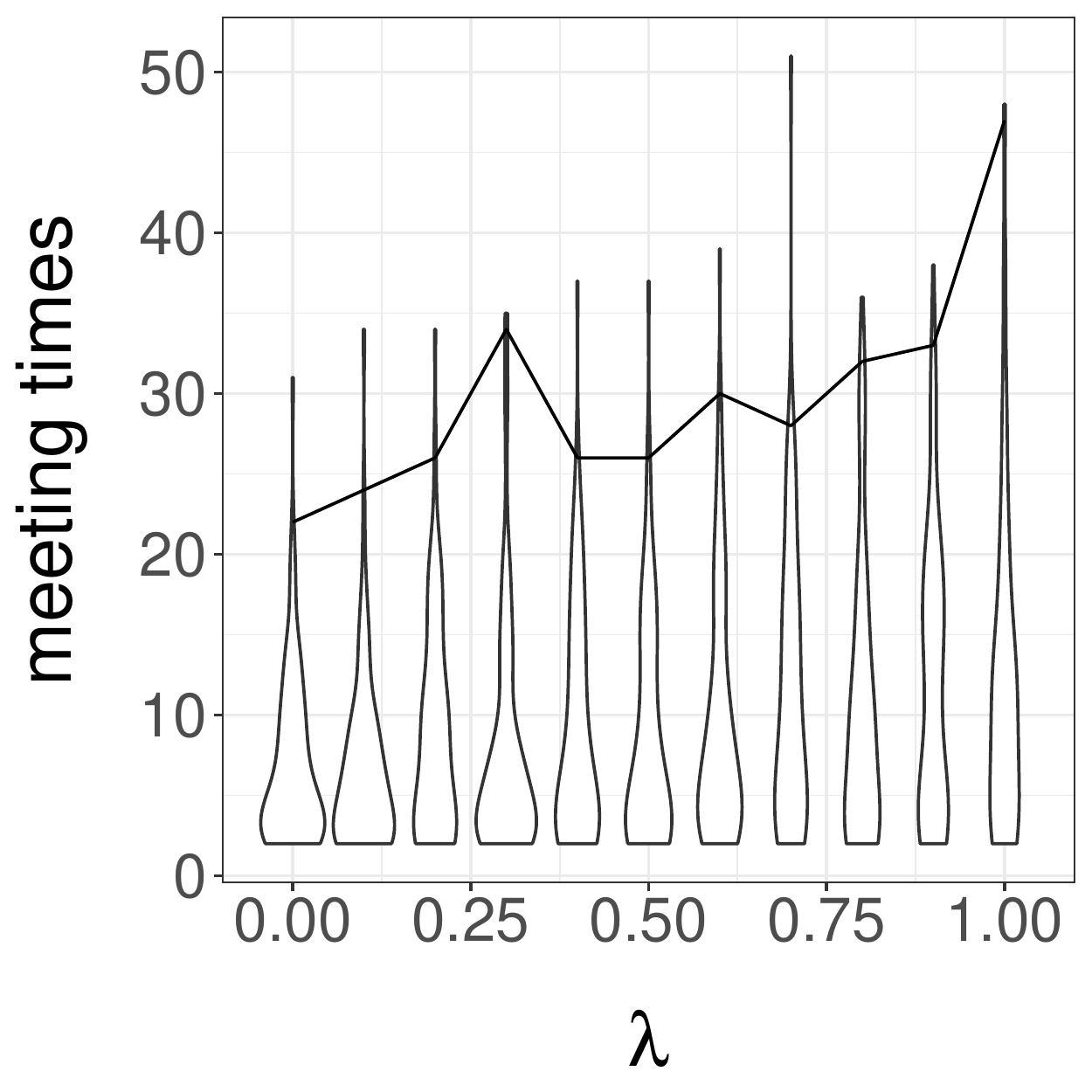}
    \caption{Meeting times for different $\pathvar$.}
    \label{fig:normal:meetings}
    \end{subfigure}
    \begin{subfigure}[t]{0.3\textwidth}
     \includegraphics[width=\textwidth]{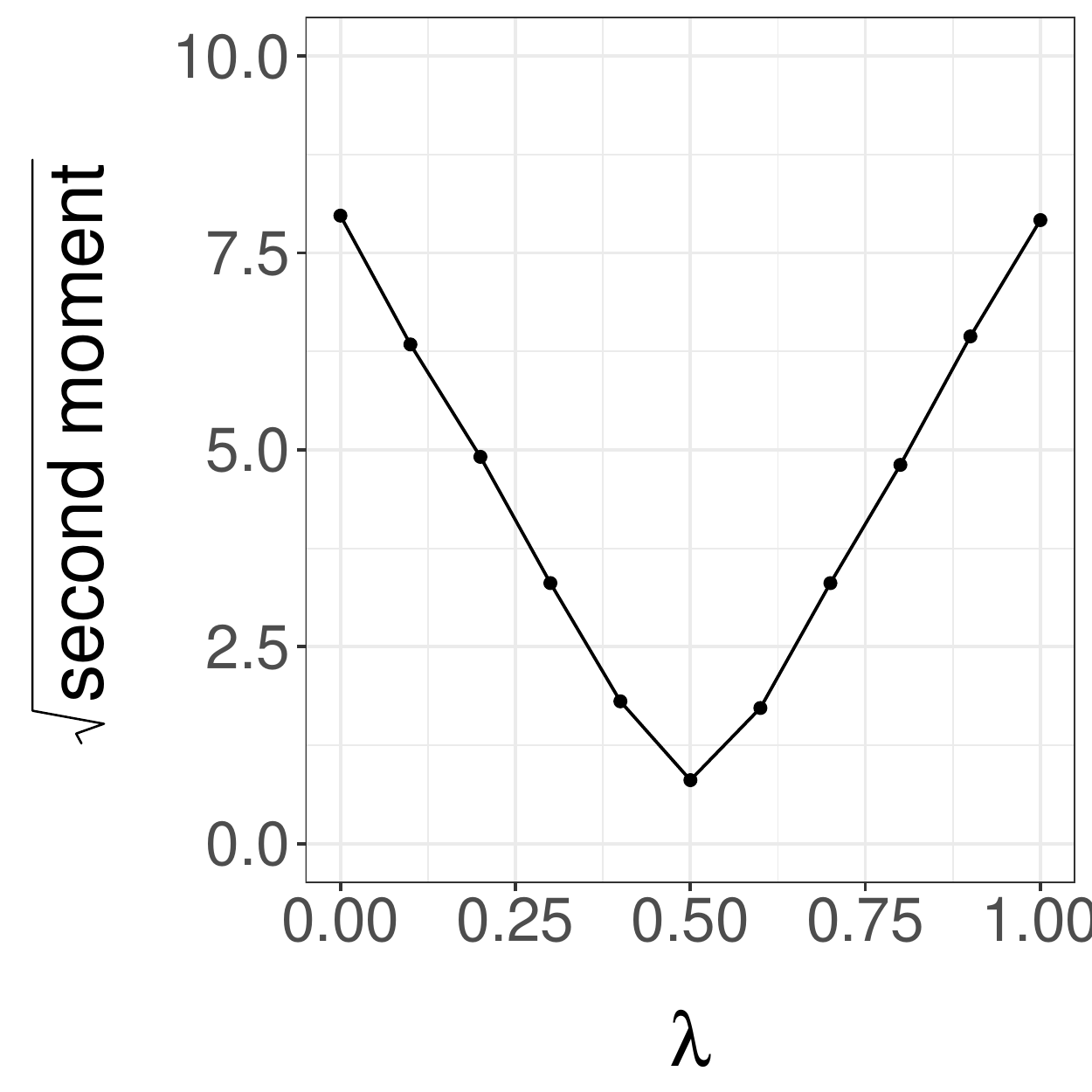}
    \caption{Estimates of $\sqrt{m_2(\pathvar)}$.}
    \label{fig:normal:sqrtmeansquare}
    \end{subfigure}
    \begin{subfigure}[t]{0.3\textwidth}
     \includegraphics[width=\textwidth]{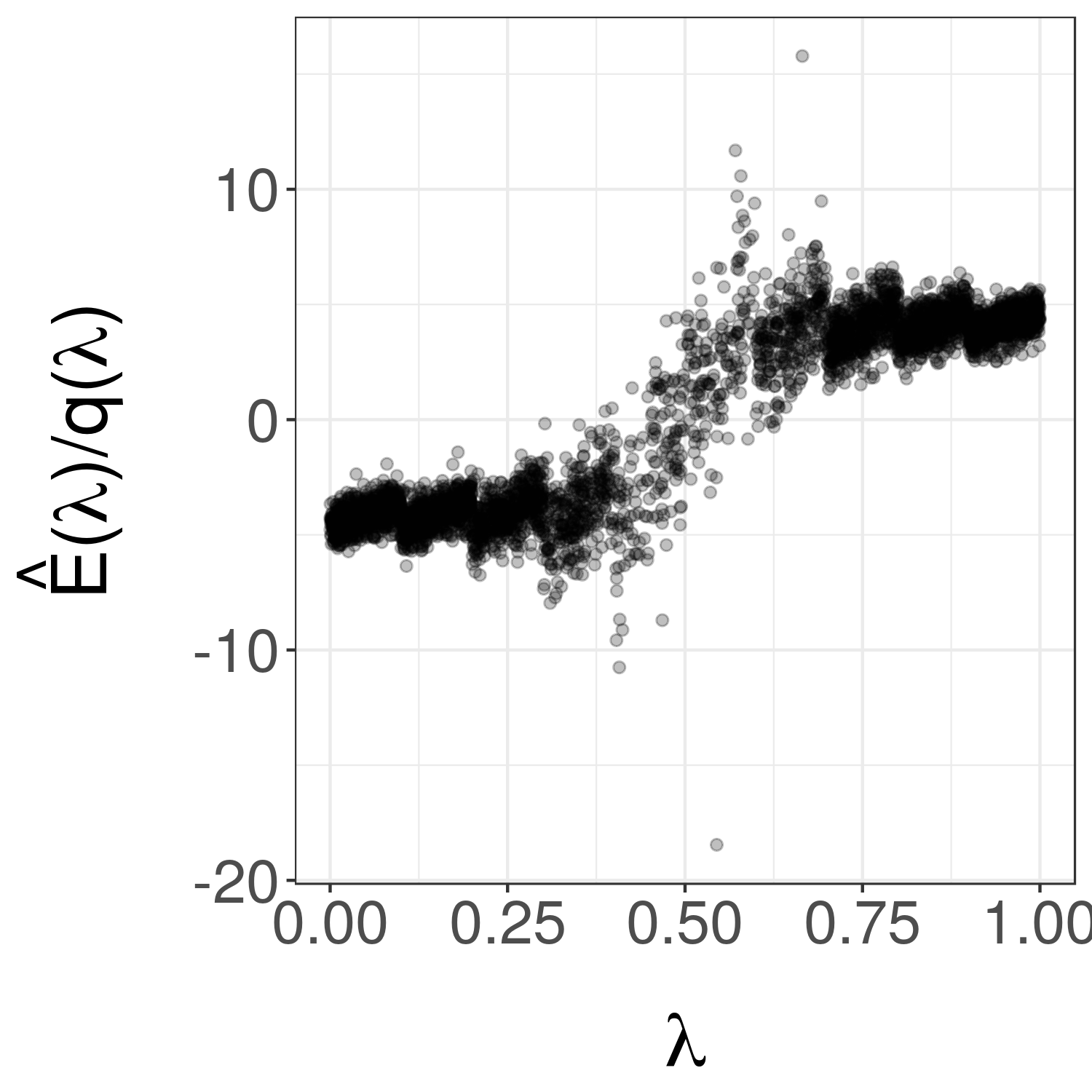}
    \caption{Estimators $\EUhat(\pathvar)/\probIS(\pathvar)$.}
    \label{fig:normal:deltaestimatorsopt}
    \end{subfigure}
\caption{\label{fig:normal:firstattempt}
Normal target example of \autoref{sec:normalexample}.
Left: distribution of meeting times
for $\pathvar \in \{0/L,1/L,\dotsc,L/L\}$ and $L=10$, in violin plots.
The $99\%$ quantiles are shown as a full line.
Middle: estimates of $\sqrt{m_2(\pathvar)}$ on a grid of values of $\pathvar$, used to define a proposal distribution $\probIS(\dif\pathvar)$.
Right: estimators $\EUhat(\pathvar)/\probIS(\pathvar)$ plotted against $\pathvar$, where $\pathvar$ is drawn from $\probIS(\dif\pathvar)$.
}
\end{figure}

We start with a grid of values of $\pathvar$: $\pathgrid{l} = l/L$ for $l \in \{0,\dotsc,L\}$,
with $L = 10$.
For each $\pathgrid{l}$, we run coupled MH chains until they meet, 100 times independently.
We obtain a distribution of meeting times $\tau$ for each $\pathvar$, represented
on \autoref{fig:normal:meetings}.
The overlaid full line represents the $99\%$ quantiles,
which we denote by $k_0,\dotsc,k_L$.
We also compute the average meeting times for each $\pathgrid{l}$,
which we denote $\bar{\tau}_0,\dotsc,\bar{\tau}_L$.
We then define
\begin{equation}
    \forall\: l \in \{0,\dotsc,L\}
    \quad
    m_{l} = 5 \times \max_{j}\del{k_j} + \max_{j}\del{\bar{\tau}_j} - \bar{\tau}_{l}
    \,.
\end{equation}
This ensures that the expected cost $C_\pathvar$, which is approximately equal to $m_\pathvar + \EMC[\tau_\pathvar]$,
is constant over $\pathvar$, while also ensuring that $m_\pathvar \geq 5 k_\pathvar$ for all $\pathvar$.

Given values of $k_l$ and $m_l$, for each $\pathgrid{l}$ in the grid of $L+1$ values defined above,
we approximate the first and second moments of $\pnorm_\pathvar$ with $100$ independent estimators.
We use these moments to redefine the initial distribution of the Markov chains, which we set to a Normal distribution
adapted to $\pnorm_\pathvar$, and to tune the proposal
standard deviation, which we set to be the estimated standard deviation of $\pnorm_\pathvar$.
At this point we could sample meeting times again and choose new values for $k_\lambda$ and $m_\lambda$, but we omit this here.
Next, we estimate $\sqrt{m_2(\pathvar)}$ for each $\pathgrid{l}$ in the grid, and define $\probIS(\dif\pathvar)$ accordingly,
following step 5 in \autoref{sec:tuning}.
The estimates of $\sqrt{m_2(\pathvar)}$ are shown in \autoref{fig:normal:sqrtmeansquare}.
This completes the tuning phase, and we can now generate unbiased estimators
$\hatratio = \EUhat(\pathvar)/\probIS(\pathvar)$ of $\logratio = \log(Z_1/Z_0)$.
We show these estimates
against $\pathvar$ in \autoref{fig:normal:deltaestimatorsopt}.
These are generated $5,000$ times
independently.
Concretely, they yield the confidence interval $[ -0.11, 0.12]$ for the estimand $\logratio=0$ at level $95\%$.

\subsubsection{Double-well example \label{sec:doublewell}}

We perform similar experiments on a path of two-dimensional distributions linking the potential
$U_0: \paramvar\mapsto (\paramvar_1+2)^2 + (\paramvar_2^2/2)$, corresponding to a Normal distribution centered at $(-2,0)$ and with diagonal variances
$(1/2, 1)$, to the potential $U_1: \paramvar\mapsto (1/10) (((\paramvar_1-1)^2-\paramvar_2^2)^2 + 10(\paramvar_1^2-5)^2+ (\paramvar_1+\paramvar_2)^4 + (\paramvar_1-\paramvar_2)^4)$.
The latter is a double-well potential, with modes around $(-2,0)$ and $(2,0)$.
By numerical integration we find $\log(Z_1/Z_0)$ to be approximately $-6.9$.
We introduce the geometric path $U_\pathvar(\paramvar) = (1-\pathvar)U_0(\paramvar) + \pathvar U_1(\paramvar)$.
For each $\pathvar$, we
start chains from a Normal centered at $(-2,-2)$ and with covariance matrix $\eye_2$, the identity matrix of size $2\times 2$.
We consider random walk MH schemes
with Normal proposal, with covariance $2 \eye_2$; 
the coupled version relies on maximal couplings of the proposals, as in the previous section.

\begin{figure}
    \centering
    \begin{subfigure}[t]{0.3\textwidth}
     \includegraphics[width=\textwidth]{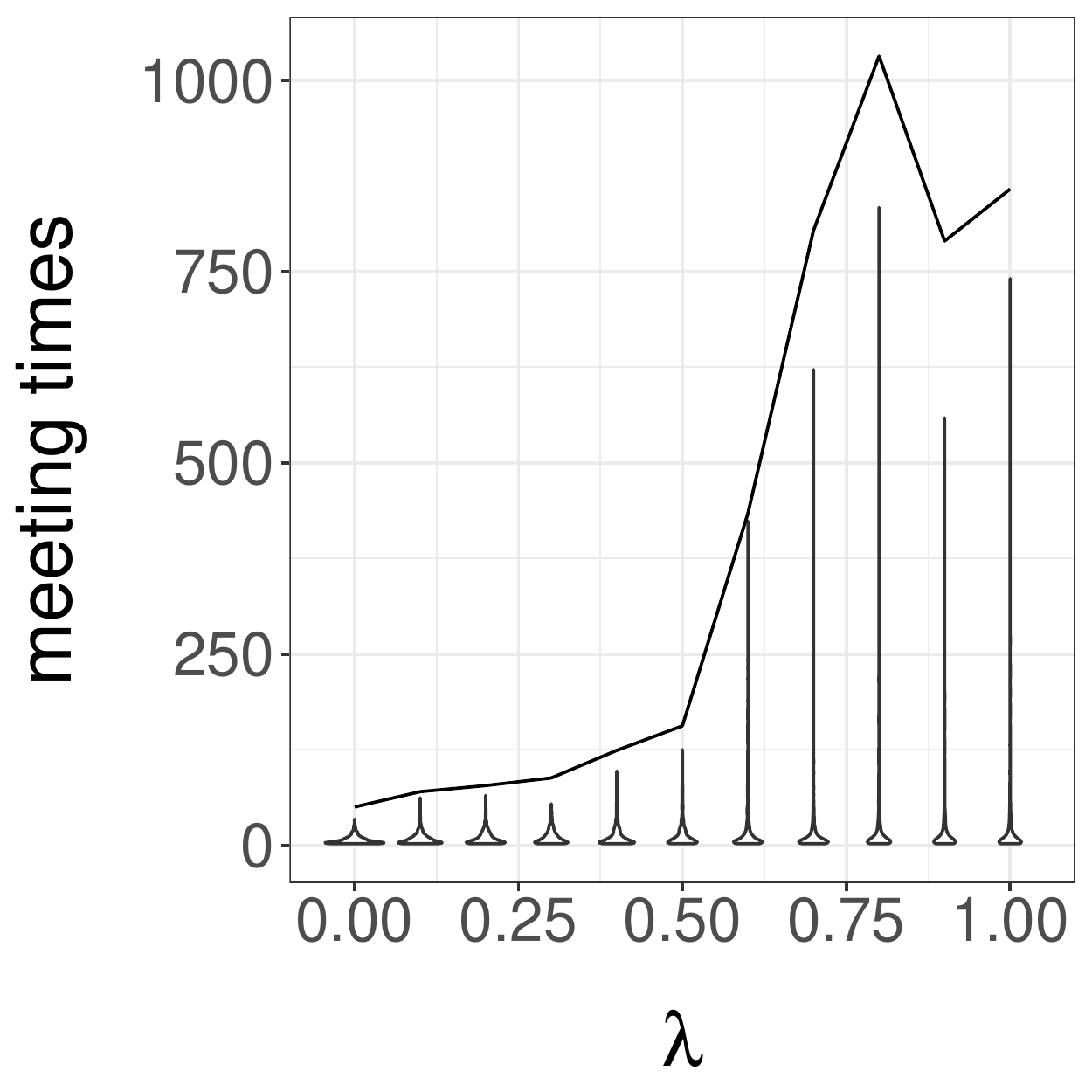}
    \caption{Meeting times for different $\pathvar$.}
    \label{fig:doublewell:meetings}
    \end{subfigure}
    \begin{subfigure}[t]{0.3\textwidth}
     \includegraphics[width=\textwidth]{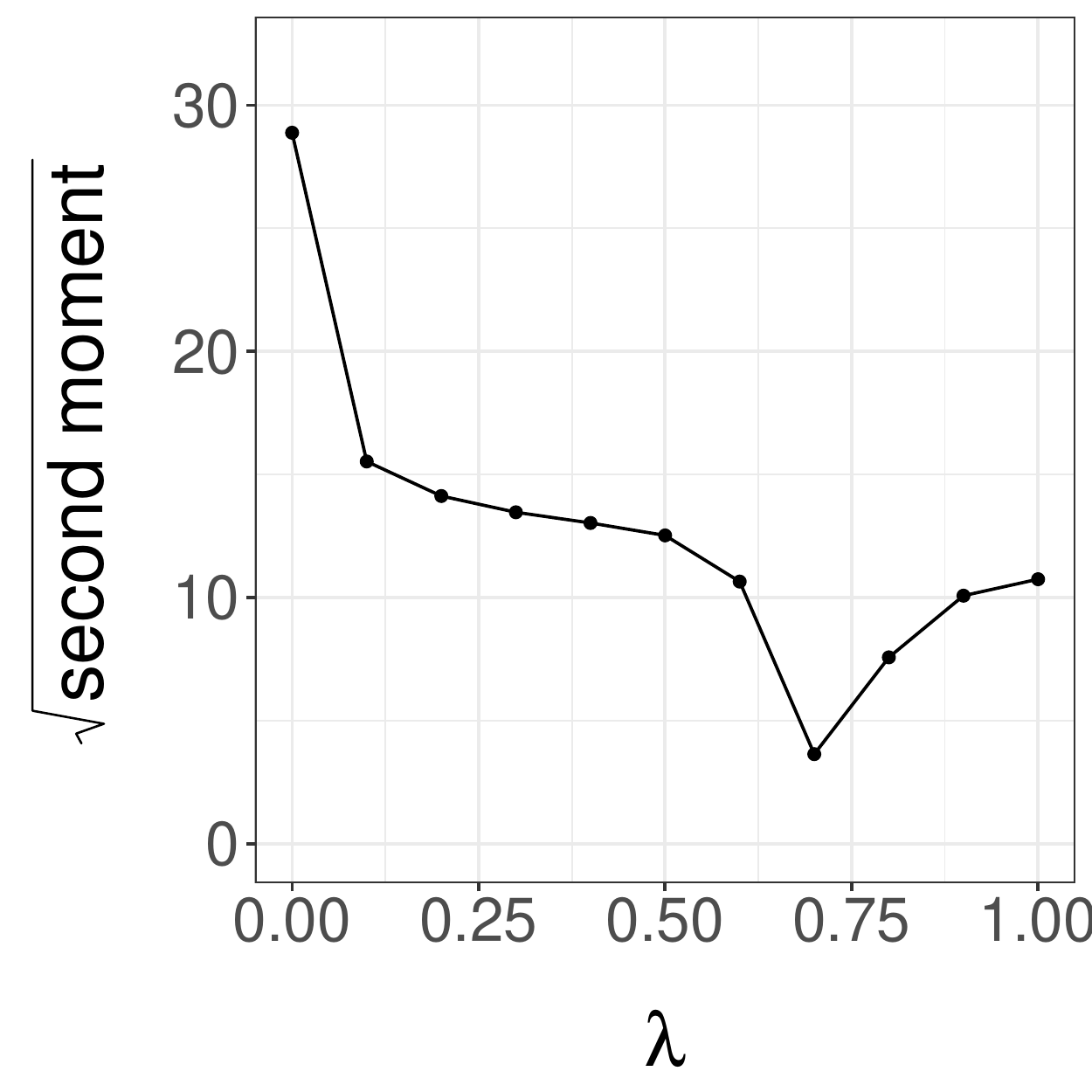}
    \caption{Estimates of $\sqrt{m_2(\pathvar)}$.}
    \label{fig:doublewell:sqrtmeansquare}
    \end{subfigure}
    \begin{subfigure}[t]{0.3\textwidth}
     \includegraphics[width=\textwidth]{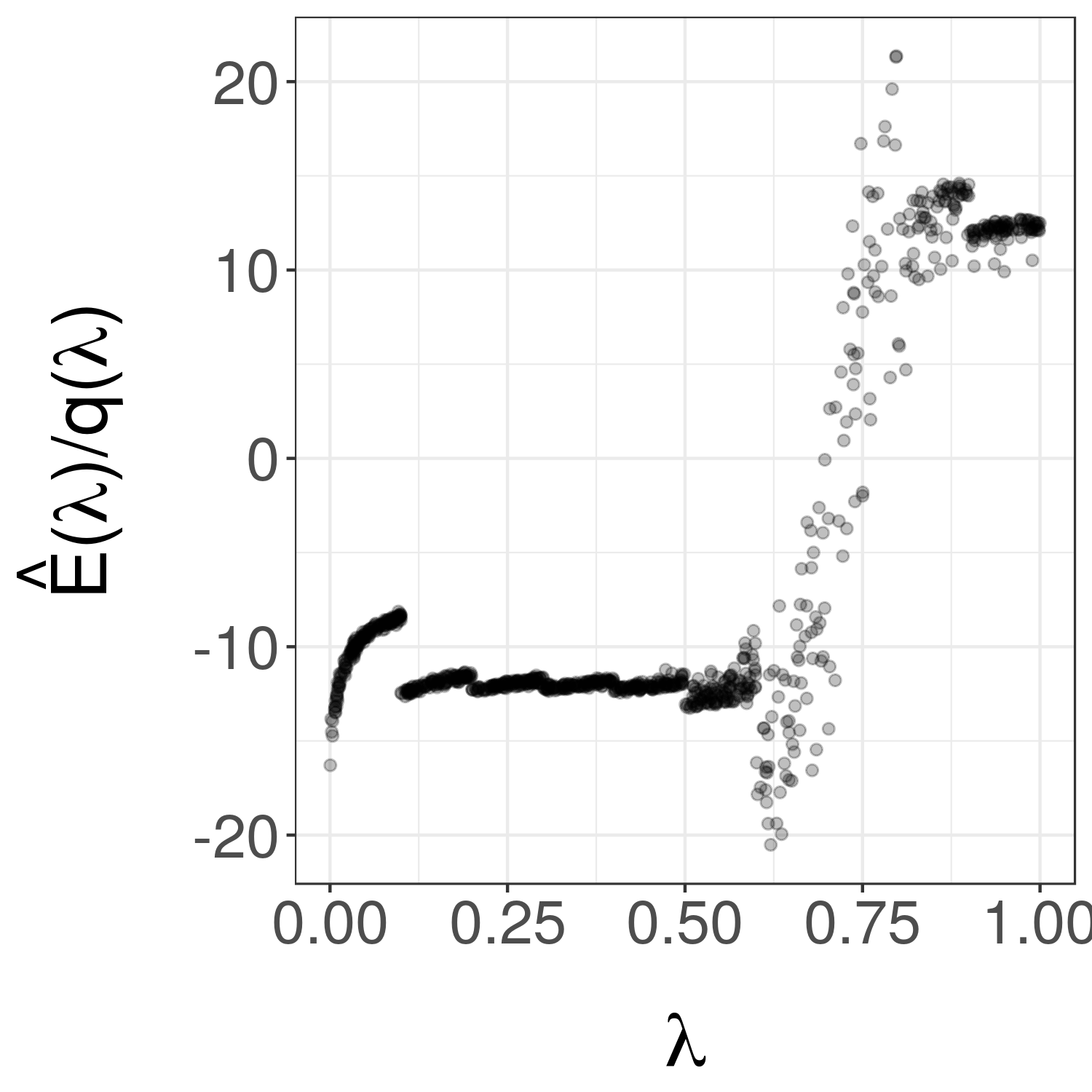}
    \caption{Estimators $\EUhat(\pathvar)/\probIS(\pathvar)$.}
    \label{fig:doublewell:deltaestimatorsopt}
    \end{subfigure}
\caption{\label{fig:doublewell:firstattempt}
Double-well potential example of \autoref{sec:doublewell}.
Left: distribution of meeting times
for $\pathvar \in \{0/L,1/L,\dotsc,L/L\}$ and $L=10$, in violin plots.
Twice the $99\%$ quantiles are shown as a full line.
Middle: estimates of $\sqrt{m_2(\pathvar)}$ on a grid of values of $\pathvar$, used to define a proposal distribution $\probIS(\dif\pathvar)$.
Right: estimators $\EUhat(\pathvar)/\probIS(\pathvar)$ plotted against $\pathvar$, where $\pathvar$ is drawn from $\probIS(\dif\pathvar)$.
}
\end{figure}

We draw $1,000$ meeting times independently, for $\pathgrid{l} = l/L$ with $l \in \{0,\dotsc,L\}$
and $L = 10$.
The distributions are shown in violin plots in \autoref{fig:doublewell:meetings}.
We observe much larger meeting times for $\pathvar$ close to one, which corresponds to the MH
chains struggling to explore both modes of the double-well potential.
We thus conservatively
set $k_l$ to be twice the $99\%$ quantiles of the meeting times, instead of the quantiles themselves.

We follow the same heuristics as in \autoref{sec:normalexample} for the choice of $m_l$.
Without modifying the initial distribution nor the proposal distribution of the MH chains,
we estimate
$\sqrt{m_2(\pathvar)}$ for each $\pathgrid{l}$ in the grid, based on 100 independent copies, and define $\probIS(\dif\pathvar)$
following again step 5 in \autoref{sec:tuning}.
The estimates of $\sqrt{m_2(\pathvar)}$ are shown in \autoref{fig:doublewell:sqrtmeansquare}.
Finally we generate unbiased estimators
$\hatratio = \EUhat(\pathvar)/\probIS(\pathvar)$ of $\logratio = \log(Z_1/Z_0)$,
and represent these estimates
against $\pathvar$ in \autoref{fig:doublewell:deltaestimatorsopt}.
These are generated $1,000$ times independently and result in the $95\%$ confidence interval $[-7.55,-6.37]$ for the estimand $\logratio \approx -6.9$.

\subsection{Logistic regression \label{sec:pgg}}

We consider a logistic regression setting, where a tuning-free Gibbs sampler can be used
to estimate $\EU(\lambda)$ conditional on each $\pathvar$, provided the path is chosen appropriately.
Let us consider the regression of $Y=(y_{1},\dotsc,y_{n})\in\left\{ 0,1\right\}^{n}$
on covariates $\Covariate=(\covariate_{1},\dotsc,\covariate_{n})\in \reals^{n \times p}$.
Throughout all the probability statements are conditioned on $\Covariate$, which we sometimes omit from the notation.
The logistic regression model assumes
$
    \prob(y_{i}=1 \mid \beta)
    =
    \expit
    \left(
        \covariate_{i}\trans
        \beta
    \right)
$,
where
$\expit: z \mapsto 1 / \del{1+\exp(-z)}$.
The parameters $\beta \in \reals^{p}$ are the regression coefficients of interest,
filling the role of the target parameters $\paramvar \in \setX$ in the presentation of \autoref{sec:methodology}.
The prior on $\beta$ is Normal $\normal(b,B)$,
with mean $b$ and covariance matrix $B$, and density denoted by $\beta \mapsto \varphi(\beta;b,B)$.
The posterior distribution has unnormalized probability density function:
\begin{equation}
    \prob(\beta \mid Y)
    \propto
    \varphi(\beta;b,B)
    \;
    \prod_{i=1}^{n} \del{
        \expit
        \del{
            \covariate_{i}\trans\beta}^{y_{i}
        }
        \del{
            1 - \expit\del{\covariate_{i}\trans\beta}
        }^{1-y_{i}}
    }
    \,.
\end{equation}
With basic manipulations this is equivalent to the following simpler form
\begin{equation}
    \prob(\beta \mid Y)
    \propto
    \varphi(\beta; b,B)
    \prod_{i=1}^n
        \frac{
            \exp(
            \covariate_i\trans \beta y_i)
        }{
            1 + \exp(\covariate_i\trans\beta)
        }
    \,.
\end{equation}

The P\'olya-Gamma Gibbs (PGG) sampler \citep{polson2013bayesian,choi2013polya} is
a Gibbs sampler that targets $\prob(\beta \mid Y)$ through the introduction of auxiliary variables $W$.
First, we recall that the P\'olya-Gamma distribution
with parameters $(1,c)$, denoted by $\text{PG}(1,c)$, has a density $x \mapsto \pg(x;c)$
defined for all $c \geq 0$, $x>0$ as
\begin{equation}
    \pg(x;c)
    = \operatorname{cosh}\del{
        \frac{c}{2}
    }
    \exp\del{-\frac{c^{2}x}{2}}
    \sum_{k=0}^{\infty}
    (-1)^{k}
    \frac{\left(2k+1\right)
    }{\sqrt{2p x^{3}}}
    \exp\del{-\frac{(2k+1)^{2}}{8x}}
    \,.
\end{equation}
Introduce $n$ auxiliary variables $W=(W_{1},\dotsc,W_{n})$,
independent of each other given $\beta$, such that $W_{i}$ follows
PG$(1,\abs{\covariate_{i}\trans\beta})$ for all $1\leq i\leq n$.
An extended target distribution is defined as $\prob(\beta,\omega \mid Y)\propto\prob(\beta \mid Y)g(\omega \mid \beta)$,
where $\omega$ denotes a realization of $W$, and $g(\omega \mid \beta)=\prod_{i=1}^{n}\pg(\omega_{i};\abs{\covariate_{i}\trans\beta})$.
The appeal of this extension is that we can write the target as
\begin{equation}
    \prob(\beta,\omega \mid Y)
    \propto
    \varphi(\beta; b,B)
    g(\omega \mid \beta)
    \prod_{i=1}^n
    \frac{\exp(\covariate_i\trans \beta y_i)}{1+\exp(\covariate_i\trans\beta)}
    \,,
\end{equation}
and therefore the conditional of $\beta$ given $\omega,Y$ simplifies to
\begin{equation}
    \prob(\beta \mid \omega,Y)
    \propto
    \varphi(\beta; b,B)
    \prod_{i=1}^n
    \exp\del{
        y_i \covariate_i\trans \beta - \frac{\covariate_i\trans\beta}{2} - \frac{(\covariate_i\trans\beta)^2 \omega_i}{2}
    }
    \,.
\end{equation}
Noting that the prior on $\beta$ is Normal, we find a Normal distribution for $\beta$ given $\omega,Y$,
with mean $\mu(\omega)$ and covariance matrix $\Sigma(\omega)$ with
$\Sigma(\omega)=(\Covariate \trans\text{diag}(\omega) \Covariate+B^{-1})^{-1},$ and $\mu(\omega)=\Sigma(\omega)(\Covariate \trans\tilde{Y}+B^{-1}b)$
where $\tilde{Y}=(y_{1}-\frac{1}{2},\dotsc,y_{n}-\frac{1}{2})$.
To summarize, the PGG sampler generates a chain $(\beta^{(t)},W^{(t)})_{t\geq0}$ in two steps:
\begin{enumerate}
    \item given $(\beta^{(t)},W^{(t)})$, draw $\beta^{(t+1)}\sim\normal\left(\mu(W^{(t)}),\Sigma(W^{(t)})\right)$,
    \item draw $W_{i}^{(t+1)}\sim\text{PG}(1,\abs{\covariate_{i}\trans\beta^{(t+1)}})$,
independently for all $i\in\left\{ 1,\dotsc,n\right\} $.
\end{enumerate}
In the experiments below, we initialize the chains from the prior distribution $\mathcal{N}(b,B)$.

\subsubsection{Normalizing constant estimation}

We first remark that the above reasoning holds when replacing the covariates $\Covariate$
by $\pathvar \Covariate$ for any $\pathvar \in [0,1]$.
This corresponds to the likelihood
$\beta \mapsto \prod_{i=1}^n \exp(\pathvar \covariate_i\trans \beta y_i)/(1+\exp(\pathvar \covariate_i\trans\beta))$,
for $\pathvar \in [0,1]$.
In the case $\pathvar = 0$,
the likelihood is equal to $2^{-n}$ for all $\beta$, while with $\pathvar = 1$, we retrieve the original likelihood.
For all $\pathvar$, we can introduce P\'olya-Gamma variables $W_{i}$ following PG$(1,\abs{\pathvar \covariate_{i}\trans\beta})$ for all $1\leq i\leq n$,
and obtain a corresponding PGG sampler.

This enables normalizing constant estimators for the logistic regression model
with little tuning, since the PGG sampler itself has no tuning parameters.
Here, for all $\pathvar,\beta$, we define
\begin{equation}
    \unnorm_\pathvar(\beta)
    =
    \varphi(\beta;b,B)
    \prod_{i=1}^n
    \frac{
        \exp(\pathvar \covariate_i\trans \beta y_i)
    }{
        1+\exp(\pathvar \covariate_i\trans\beta)
    }
    \,,
\end{equation}
so that,
\begin{equation}
    \log \unnorm_\pathvar(\beta)
    = \log \varphi(\beta;b,B)
    + \sum_{i=1}^n \cbr{
        \pathvar \covariate_i\trans \beta y_i
        - \log\del{
            1+\exp(\pathvar \covariate_i\trans\beta)
        }
    }
    \,,
\end{equation}
and thus
\begin{equation}
    \nabla_\pathvar \log \unnorm_\pathvar(\beta)
    =
    \sum_{i=1}^n \cbr{
        \covariate_i\trans \beta y_i
        - \frac{
            \covariate_i\trans\beta\exp(\pathvar \covariate_i\trans\beta)
        }{
            1+\exp(\pathvar \covariate_i\trans \beta)
        }
    }
    \,,
\end{equation}
which can be used to carry out the path sampling calculations.
Note also that $Z_0 = 2^{-n}$, so the proposed estimator $\hatratio$ will have
expectation $\log Z_1 + n \log 2$.
Finally, notice that the function $\beta \mapsto \nabla_\pathvar \log \unnorm_\pathvar(\beta)$
is linear in $\beta$ far away from $\beta=0$, while $\pnorm_\pathvar$ has at most the Gaussian tails of the prior $\normal(b,B)$ (by crudely upper-bounding the likelihood by a constant).
Therefore, any power of $\nabla_\pathvar \log \unnorm_\pathvar(\beta)$ has finite expectation under $\pnorm_\pathvar(\dif\beta)$, for any $\pathvar$, and thus we can check that path sampling estimators have a finite variance.

\begin{figure}
    \centering
    \begin{subfigure}[t]{0.3\textwidth}
     \includegraphics[width=\textwidth]{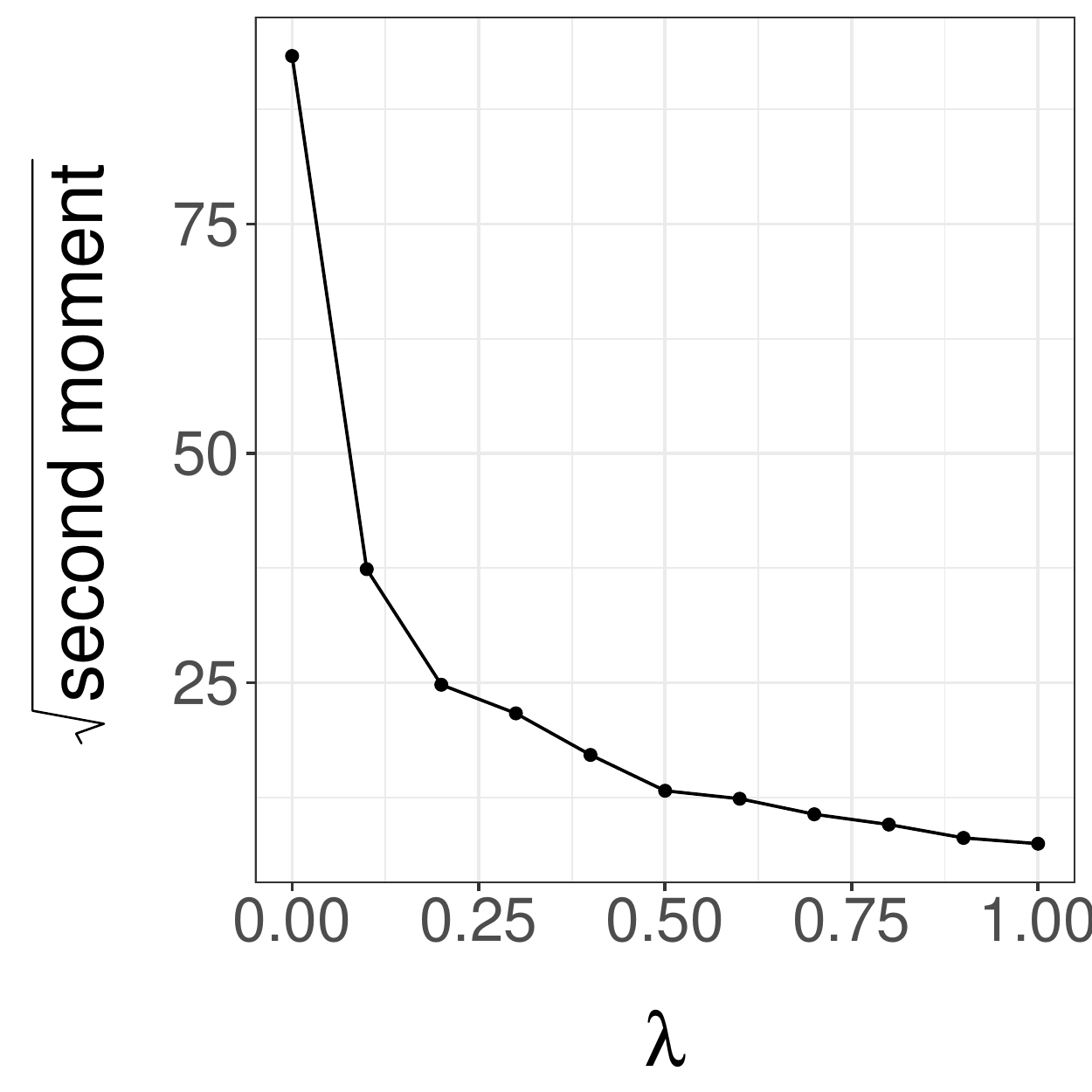}
    \caption{Estimates of $\sqrt{m_2(\pathvar)}$.}
    \label{fig:pgg:sqrtmeansquare}
    \end{subfigure}
    \begin{subfigure}[t]{0.3\textwidth}
     \includegraphics[width=\textwidth]{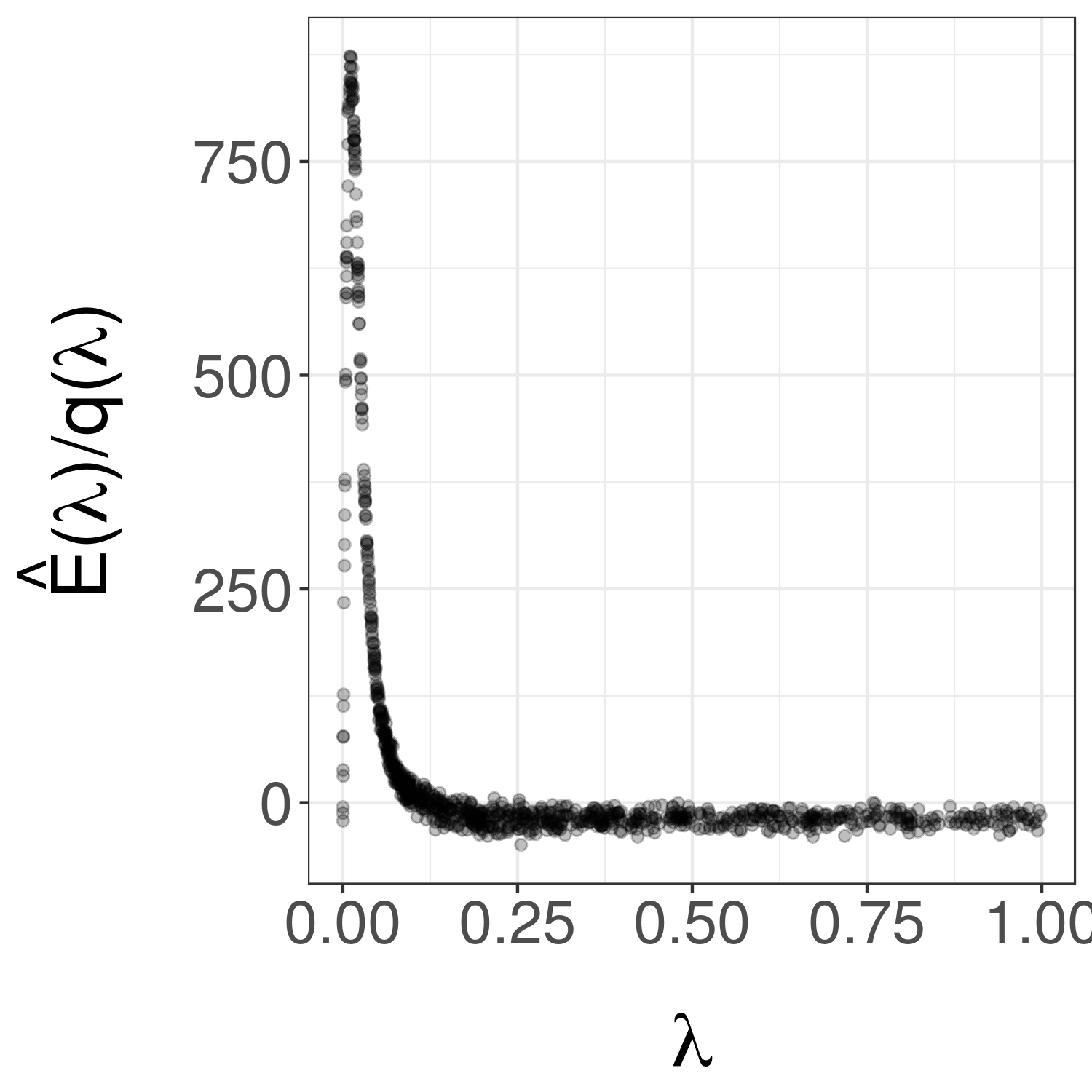}
    \caption{Estimators $\EUhat(\pathvar)/\probIS(\pathvar)$.}
    \label{fig:pgg:deltaestimatorsopt}
    \end{subfigure}
    \begin{subfigure}[t]{0.3\textwidth}
     \includegraphics[width=\textwidth]{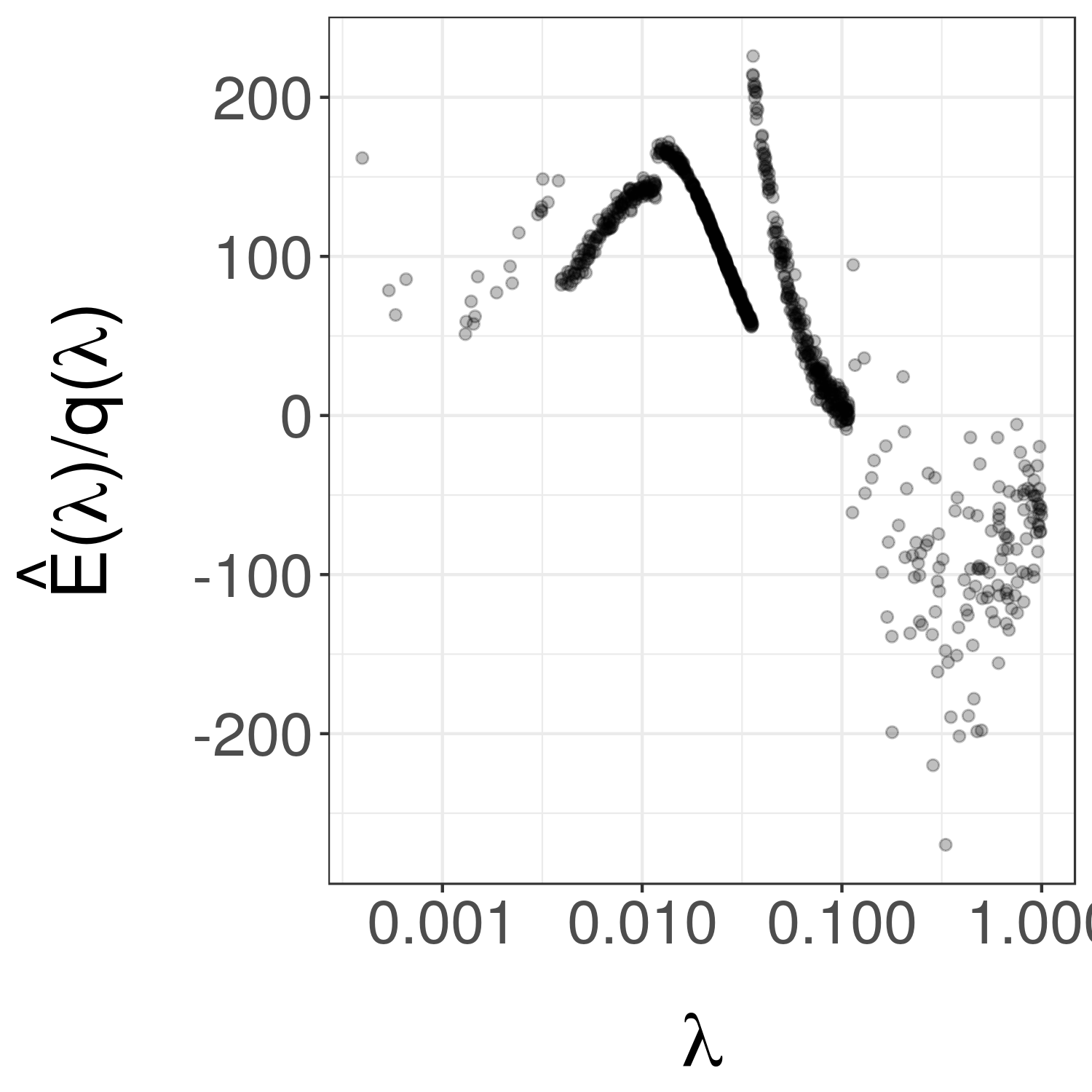}
    \caption{Estimators $\EUhat(\pathvar)/\probIS(\pathvar)$, with a better choice of $\probIS(\pathvar)$.}
    \label{fig:pgg:deltaestimatorsopt:newgrid}
    \end{subfigure}
\caption{\label{fig:pgg:firstattempt}
Logistic regression example of \autoref{sec:pgg}.
Left: estimates of $\sqrt{m_2(\pathvar)}$ on an equispaced grid of values of $\pathvar$, used to define a proposal distribution $\probIS(\dif\pathvar)$.
Middle: estimators $\EUhat(\pathvar)/\probIS(\pathvar)$ plotted against $\pathvar$, where $\pathvar$ is drawn from $\probIS(\dif\pathvar)$, designed on an equispaced grid of values of $\pathvar$ in $[0,1]$.
Right: estimators $\EUhat(\pathvar)/\probIS(\pathvar)$ plotted against $\pathvar$, where $\probIS(\dif\pathvar)$ is designed based on values of $\pathvar$ equispaced on the logarithmic scale.
}
\end{figure}

We consider a synthetic data set with $n=1000$ rows and $p=7$ columns.
The covariates are generated from a standard Normal distribution
and the outcome is generated from the model with $\beta^\star = (0,0.1,0.2,0.3,0.4,0.5,0.6)$.
The prior mean $b$ is set to zero
and the covariance $B$ to a diagonal matrix with entries equal to $10$.
We start by gridding the interval $[0,1]$, and for each value $\pathgrid{l} =
l/L$ with $L=10$, we set $k$ as the $99\%$ quantile of the meeting times for
the coupled PGG sampler, based on 1000 independent runs.
We set $m$ as in the previous sections,
to make the average cost approximately constant over $\pathvar$.
Next we estimate the second moments
of $\EUhat(\pathvar)$ on the grid of values of $\pathvar$, we design a proposal $\probIS(\dif\pathvar)$
following step 5 in \autoref{sec:tuning}, and obtain the estimates of \autoref{fig:pgg:sqrtmeansquare}.

We obtain the $1,000$ independent estimators $\EUhat(\pathvar)/\probIS(\pathvar)$ shown in \autoref{fig:pgg:deltaestimatorsopt},
leading to a $95\%$ confidence interval of $[63, 89]$ on $\logratio=\log(Z_1/Z_0)$.
The actual value is found to be close to $70$ using importance sampling based on a Laplace approximation to the posterior,
accurate in the present example (see below, and also related discussions in \citet{bardenet2017markov}).
We can see from \autoref{fig:pgg:deltaestimatorsopt} that the estimates
take very large values for $\pathvar$ close to zero.
This suggests that, instead of choosing an equispaced grid of values
of $\pathvar$ on $[0,1]$ when designing $\probIS(\dif\pathvar)$, we could aim at a higher resolution towards the left end of the interval $[0,1]$.

Therefore we consider a grid of values of $\pathvar$ equispaced on the logarithmic scale:
$\pathgrid{l} = \exp(-L+l)$ for $l=0,\dotsc,L$ with $L=10$.
Going through the exact same tuning steps, we obtain the $1,000$ estimators
of \autoref{fig:pgg:deltaestimatorsopt:newgrid}, leading to the narrower confidence interval
$[64.7, 74.7]$ at level $95\%$ (with a width of 10 instead of 36 for the previous one).
This illustrates the potential gains obtained by carefully choosing the distribution $\probIS(\dif\pathvar)$.

We conclude this section by noting that more dramatic gains can be obtained by
changing the path of distributions.
In the context of logistic regression with
$n\gg p$, the Laplace approximation of the posterior, defined
as $\normal(\hat{\beta}_\text{MLE}, \hat{V})$ where
$\hat{\beta}_{\text{MLE}}$ is the maximum likelihood estimator and $\hat{V}$ is
the inverse of minus the Hessian of the log-likelihood evaluated at
$\hat{\beta}_{\text{MLE}}$, seems to be very accurate.
We thus introduce a geometric
path $(\pnorm_\pathvar)$ between the Laplace approximation and the posterior
distribution.
We use a random walk MH algorithm to target $\pnorm_\pathvar$ for all
$\pathvar\in[0,1]$, with proposal covariance matrix equal to $\hat{V}/p$ where
the dimension $p$ is equal to $7$.
To couple the MH algorithms, we use 
strategy that combines reflection and maximal couplings, as described in \citet{jacob2017unbiased}.
The initial distribution of the chains is chosen to be the Laplace approximation.
For $\pathvar = 0$, we obtain $k=127$ as the $99\%$ quantile of the meeting times,
and we set $m = 5k$.
We use these values of $k$ and $m$ for all $\pathvar$, and
we choose $\probIS(\dif\pathvar)$ to be uniform on $[0,1]$.
With $100$ independent
estimators we obtain a confidence interval of $[70.24, 70.26]$ at $95\%$ for
$\log Z_1 + n \log 2$.
This is orders of magnitude narrower than the
previous intervals, for a smaller computational cost.
The choice
of paths can thus play a critical role in the efficiency of the proposed estimators,
and approximations of the posterior distribution can be used to construct such paths.

\subsubsection{Cross-validation}

We now consider the approximation of CV in \autoref{eq:cvobjective}.
We consider a leave-one-out criterion, with $n_T = n-1$ and $n_V= 1$.
We thus construct paths between the posterior given the training data $T$,
with normalizing constant $\prob(T)$, and the posterior given all the data $(T,V)$,
with normalizing constant $\prob(T,V)$.

Our first path follows the reasoning of the previous section:
we can multiply the covariates in the validation set by $\pathvar \in [0,1]$
to preserve the original structure of the likelihood
and thus to enable a similar PGG sampler.
The unnormalized densities are then
\begin{equation}
    \forall \: \pathvar \quad \forall\:\beta \quad \unnorm_\pathvar(\beta) = \varphi(\beta; b, B)
        \left\{\prod_{\covariate,y \in T} \frac{\exp(\covariate\trans \beta y)}{1 + \exp(\covariate\trans \beta)}\right\}
        \left\{\prod_{\covariate,y \in V} \frac{\exp(\pathvar \covariate\trans \beta y)}{1 + \exp(\pathvar \covariate\trans \beta)}\right\} 
    \,.
\end{equation}
Note that $Z_0$ is here equal to $-n_V \log(2) + \log \prob(T)$, and that the derivative of $\log \unnorm_\pathvar(\beta)$ is easily computed as
\begin{equation}
    \nabla_\pathvar \log \unnorm_\pathvar(\beta)
    =
    \sum_{\covariate,y \in V}
    \cbr{
        \covariate\trans \beta y - \frac{\covariate\trans\beta\exp(\pathvar \covariate\trans\beta)}{1+\exp(\pathvar \covariate\trans \beta)}
    }
    \,.
    \end{equation}
Again we see that this is essentially a linear function of $\beta$ and thus its moments under $\pnorm_\pathvar$ are finite for all $\lambda$.

\begin{figure}
    \centering
    \begin{subfigure}[t]{0.3\textwidth}
     \includegraphics[width=\textwidth]{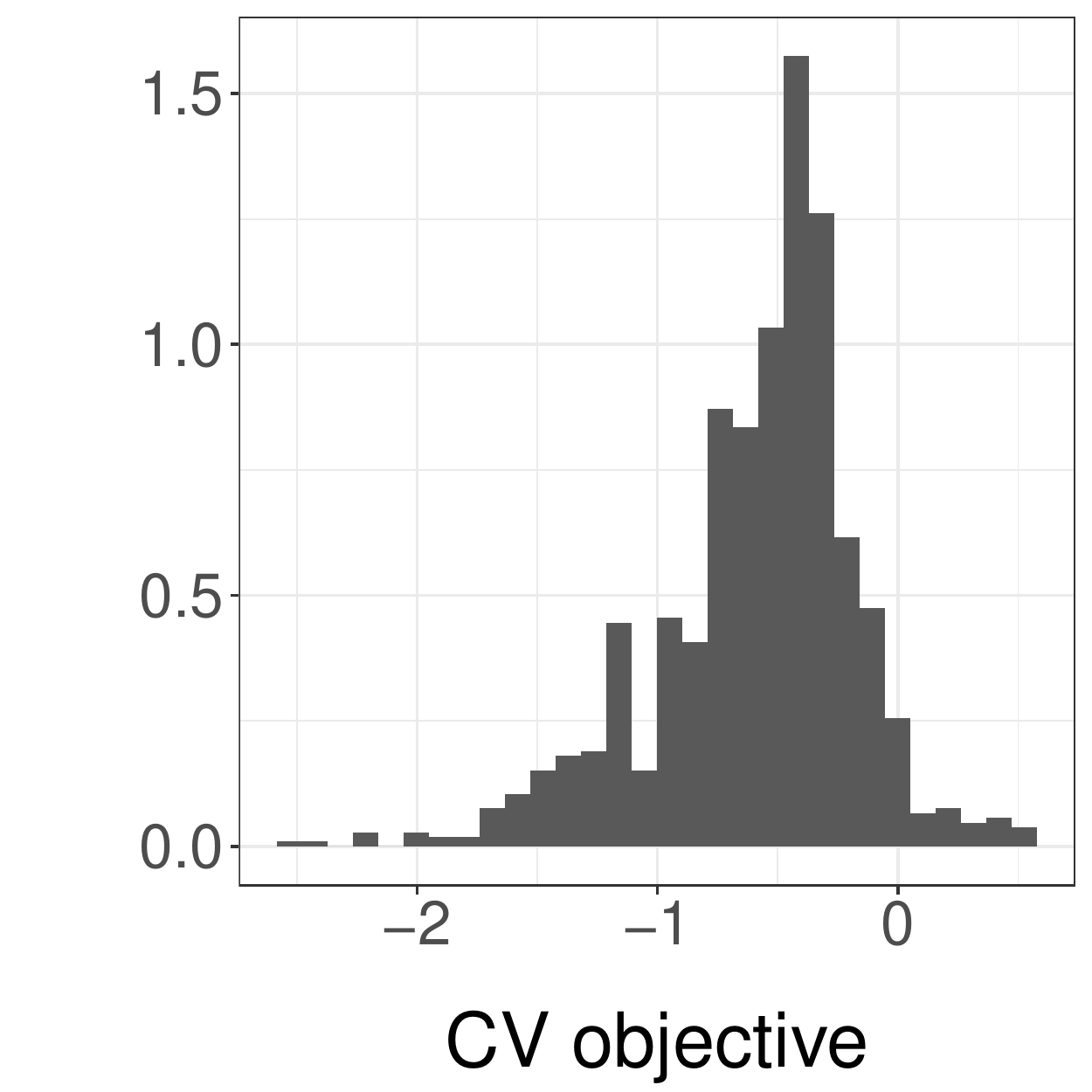}
    \caption{With PGG.}
    \label{fig:logistic:pgg:cv}
    \end{subfigure}
    \hspace*{1cm}
    \begin{subfigure}[t]{0.3\textwidth}
     \includegraphics[width=\textwidth]{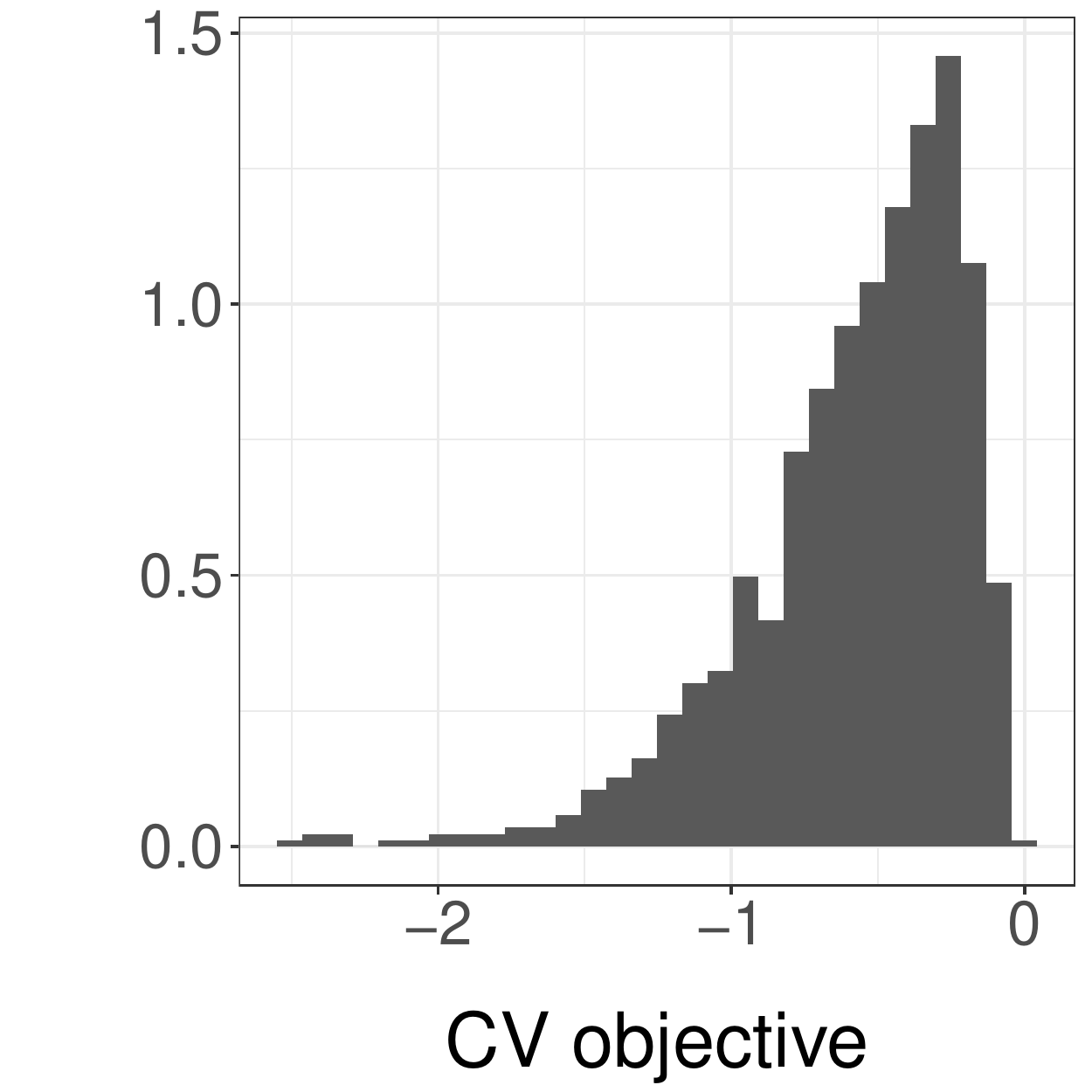}
    \caption{With MH on tempered posteriors.}
    \label{fig:logistic:otherpath:cv}
    \end{subfigure}
    \caption{\label{fig:logistic:cv} Cross-validation (leave-one-out) objective of \autoref{eq:cvobjective} for the logistic regression of \autoref{sec:pgg}.
    Left: unbiased estimators of CV obtained with coupled PGG samplers, $1000$ times independently.
    Right: unbiased estimators of CV obtained with a path linking partial posteriors to the full posterior by tempering,
    and using coupled random walk MH.
}
\end{figure}

To tune the procedure, we obtain meeting times for the coupled PGG sampler based on the full data set,
and choose $k$ as a $99\%$ quantile (here equal to $8$), and $m=5k =40$.
Recall that the PGG sampler itself has no tuning parameters.
Then, drawing a validation set at random $1,000$ time independently, generating $\pathvar$ uniformly on $[0,1]$ and obtaining the associated
estimator $\EUhat(\pathvar)$, we obtain unbiased estimators of CV in \autoref{eq:cvobjective}.
We plot a histogram of these estimators in \autoref{fig:logistic:pgg:cv}.
A $95\%$ confidence interval for the CV objective is obtained as $[-0.62, -0.56]$.

Alternatively, we introduce a geometric path between the posterior given $T$ and given $T,V$,
which corresponds to the unnormalized densities
\begin{equation}\forall \: \pathvar \quad \forall \: \beta \quad \unnorm_\pathvar (\beta) = \left\{\prod_{\covariate,y \in T}\frac{\exp(\covariate\trans \beta y)}{1+\exp(\covariate\trans\beta)} \right\}
    \left\{\prod_{\covariate,y \in V}\frac{\exp(\covariate\trans \beta y)}{1+\exp(\covariate\trans\beta)} \right\}^\pathvar \; \varphi(\beta; b,B),
\end{equation}
with associated gradient of logarithm,
\begin{equation}
    \nabla_\pathvar \log \unnorm_\pathvar(\beta) = \sum_{\covariate,y \in V}
    \cbr{
        \covariate\trans \beta y
        - \log \del{1+\exp(\covariate\trans \beta)}
    }
    \,.
\end{equation}
As with the previous path, we can check that powers of $\nabla_\pathvar \log \unnorm_\pathvar(\beta)$ have finite expectation under $\pnorm_\pathvar$
for all $\pathvar$.

For this path, we use random walk MH as in the previous section, with initial distribution and proposal covariance tuned using a Laplace
approximation of the posterior distribution.
We obtain a $99\%$ quantile of meetings at $k=135$ and set $m=5k$.
Over $1,000$ independent experiments we obtain unbiased estimators of the CV objective shown in \autoref{fig:logistic:otherpath:cv}.
The associated $95\%$ confidence interval for CV is $[-0.60, -0.56]$.
Thus, this second approach appears to be
marginally more efficient than the first one; the cost comparison is made slightly difficult by the fact that PGG and MH 
have different costs per iteration.

\subsubsection{Leukemia survival data \label{sec:leukemia}}

We follow \citet{vehtari2017practical} and consider the leukemia data presented in
\citet{feigl1965estimation} and used as illustration in \citet{epifani2008case}.
We use the data formatted as in the package \texttt{BGPhazard}, see \citet{garcia2016introduction}.
The outcome is taken to be one if the survival time (column \texttt{time} of \texttt{leukemiaFZ}) is larger or equal to $50$ weeks,
zero otherwise, and the two covariates are the columns \texttt{wbc} and \texttt{AG}, corresponding to counts of white blood cells and the outcome of
a test related to white blood cell characteristics.
There are 31 patients in the sample, so $n = 31$, and we consider leave-one-out cross-validation,
i.e. $n_T = n-1$ and $n_V = 1$.

\begin{figure}
    \centering
    \begin{subfigure}[t]{0.3\textwidth}
     \includegraphics[width=\textwidth]{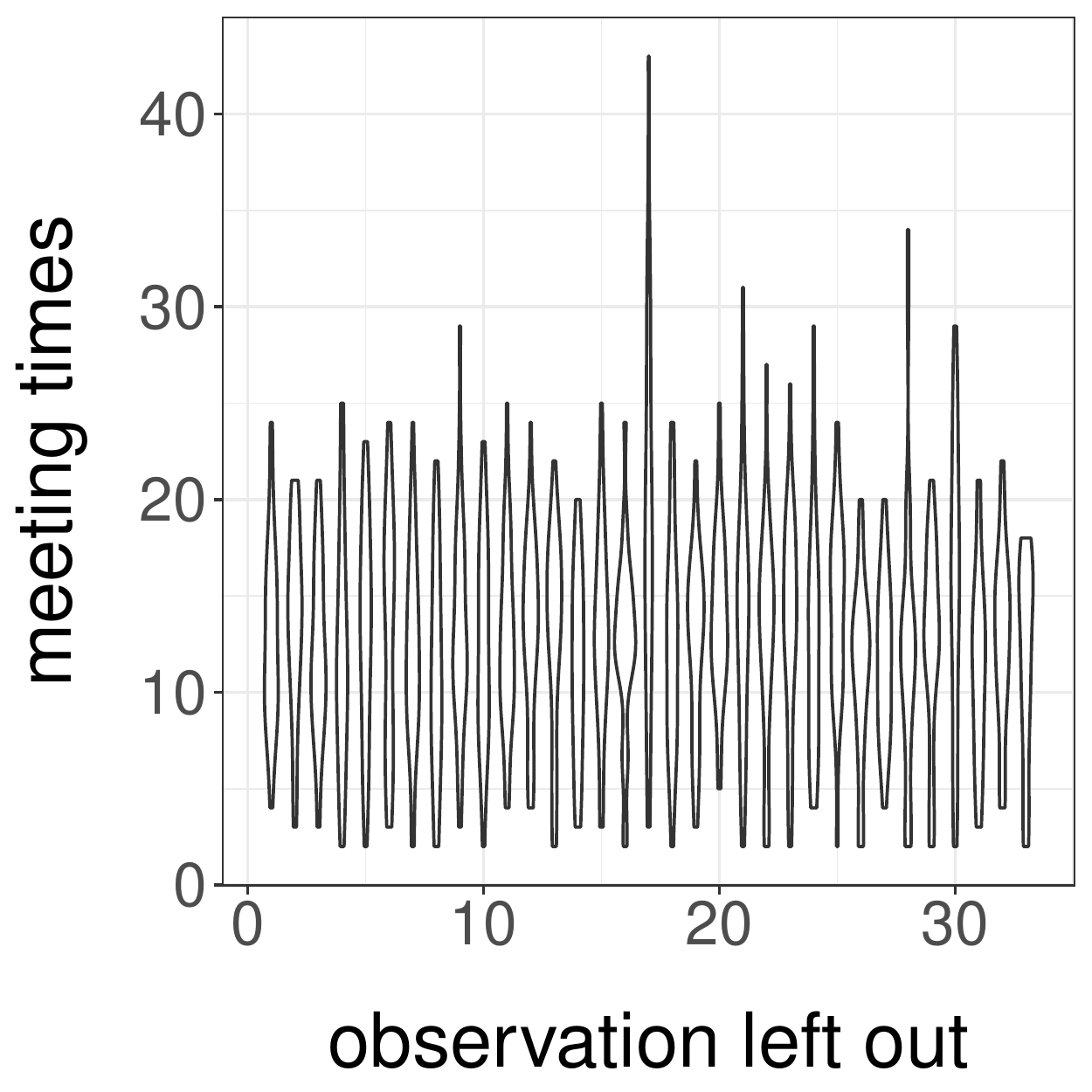}
    \caption{Meeting times  against index of left-out observation.}
    \label{fig:leukemia:meetings}
    \end{subfigure}
    \begin{subfigure}[t]{0.3\textwidth}
     \includegraphics[width=\textwidth]{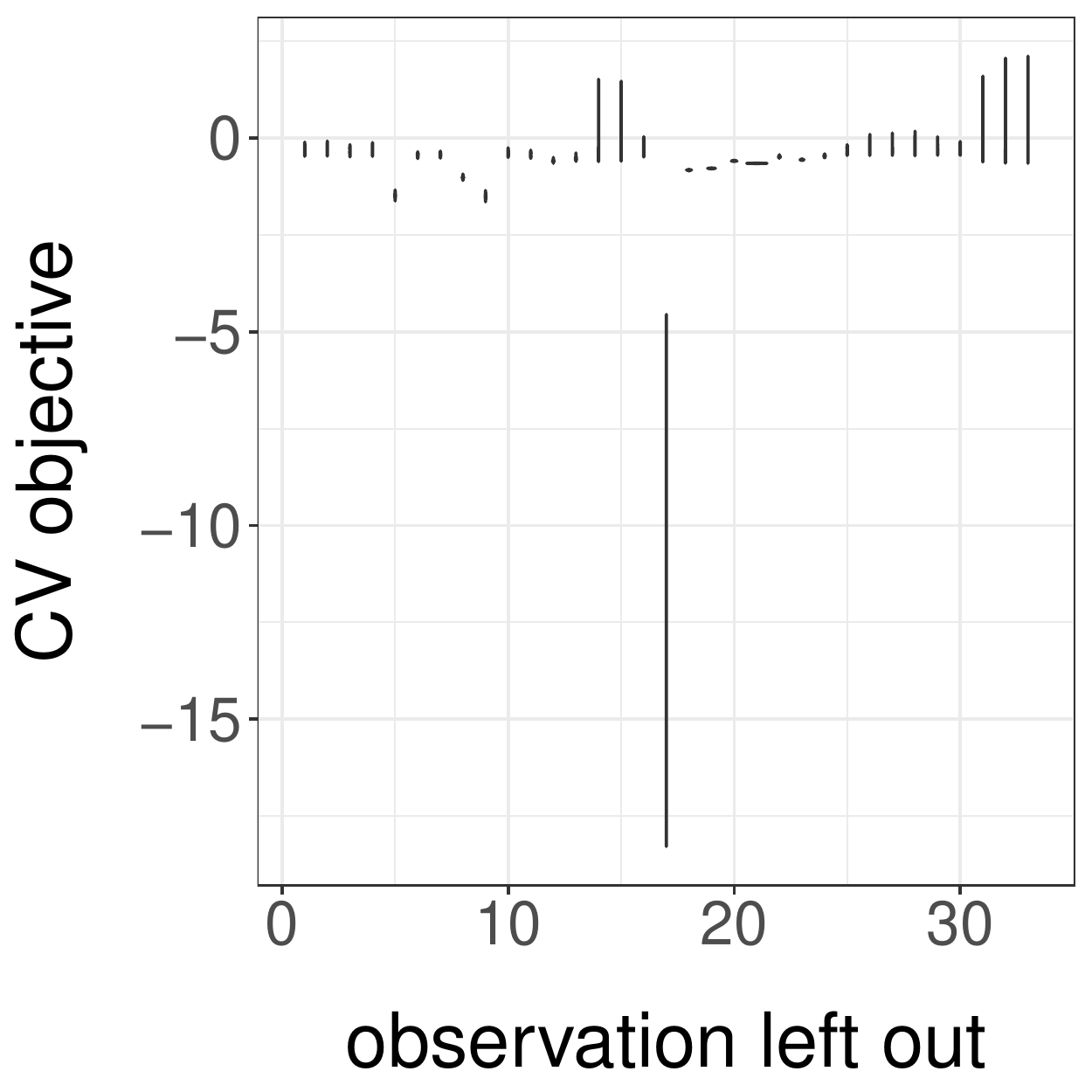}
    \caption{CV objective against index of left-out observation.}
    \label{fig:leukemia:cvindex}
    \end{subfigure}
    \begin{subfigure}[t]{0.3\textwidth}
     \includegraphics[width=\textwidth]{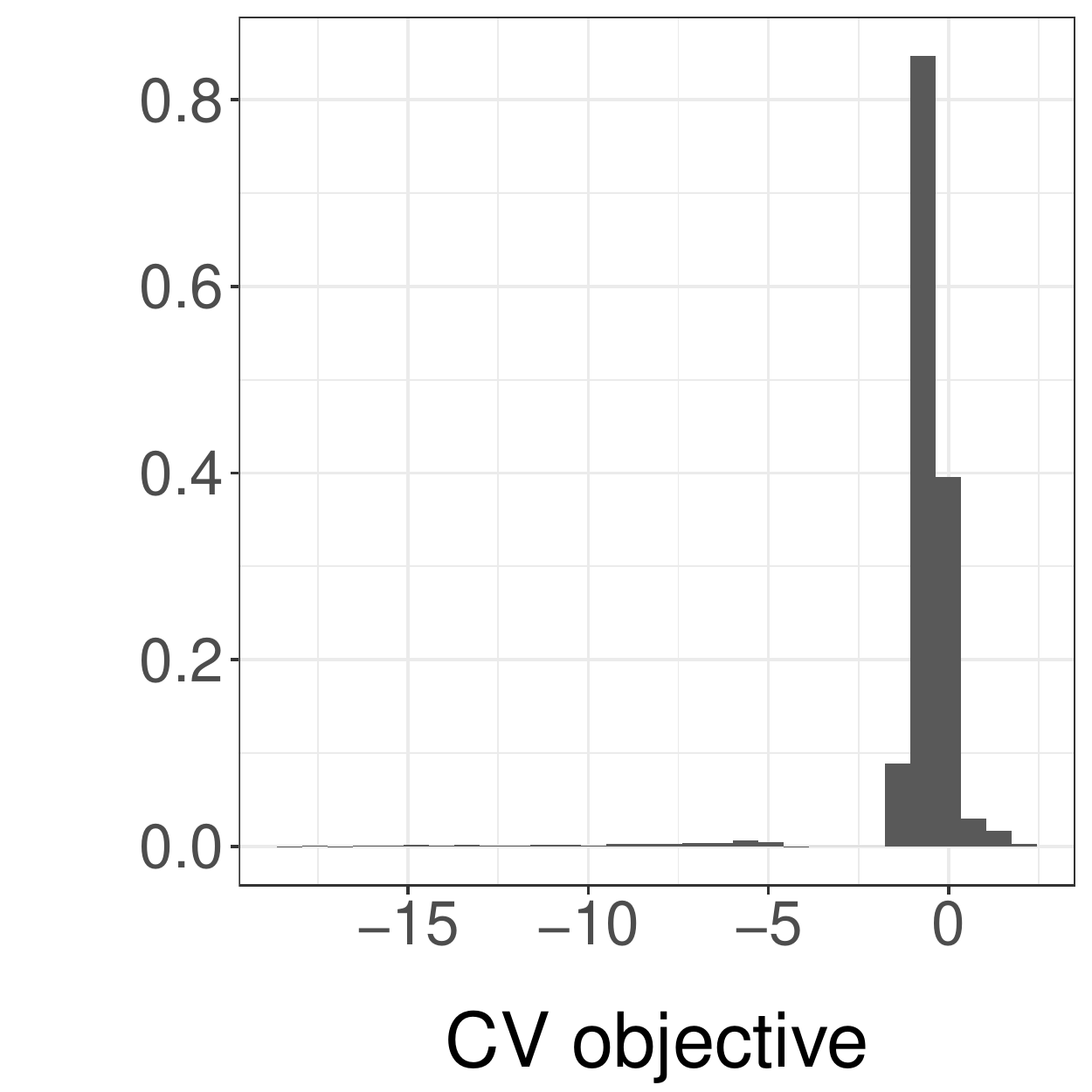}
    \caption{Histogram of CV estimates.}
    \label{fig:leukemia:cvhist}
    \end{subfigure}
    \caption{\label{fig:leukemia} Cross-validation (leave-one-out) objective of \autoref{eq:cvobjective} for the logistic regression on
    the leukemia data of \autoref{sec:leukemia}, using coupled PGG samplers. Left: meeting times against the index of left-out observation.
    Middle: CV objective against the index of left-out observation. Right: Histogram of all the $10,000$ CV estimates combined.
}
\end{figure}

We introduce a path of distributions amenable to PGG sampling, as in the previous sections.
Sampling uniformly the index of the observation to be left out, then sampling $\pathvar$ uniformly in $[0,1]$,
and finally running coupled PGG chains targeting $\pnorm_\pathvar$, we record the meeting times.
We do so 1,000 times
independently, and show the results as a function of the index of the observation left out in \autoref{fig:leukemia:meetings}.

Based on this plot we select $k=100$, conservatively, and $m=5k = 500$ for all runs.
We then generate $10,000$ unbiased estimators of CV.
We plot the estimators against the index of the left-out observation in \autoref{fig:leukemia:cvindex}, and we note that the values
are very different for one particular index, here equal to 17.
In \autoref{fig:leukemia:cvhist} we plot a histogram of the estimates of the CV objective,
putting all the indices together.
From these estimates we obtain a $95\%$ confidence interval $[-0.72, -0.66]$ for the leave-one-out CV objective.
Thus we see that the proposed estimators can have a larger variance for certain splits of the data compared to others.
Investigating further
the behavior of the estimators for certain splits, one might be able to reduce the variance, for instance by tuning the proposal distribution $\probIS(\dif\pathvar)$,
or by changing the path.
Our estimators of CV might also be considered satisfactory as they stand. In 
any case, they do not suffer from infinite variance issues
typically associated with importance sampling, when using a proposal distribution that has lighter tails than the target distribution.

\subsection{Linear regressions \label{sec:linearregression}}

We next consider linear regressions, which have been used to illustrate
Bayesian cross-validation e.g. in
\citet{alqallaf2001cross,peruggia1997variability,vehtari2017practical}.

\subsubsection{Mammal weight data \label{sec:mammal}}

The first example is taken from \cite{alqallaf2001cross}.
The data comprise of $n=62$ observations, each corresponding to an animal (arctic fox, owl monkey, etc).
For each animal, the data set contains the body weight and the brain weight.
The covariate $\covariate_i$ of animal $i$ is a vector, with first entry equal to $1$  and second entry equal to the logarithm of body weight,
while the outcome $y_i$ is the logarithm of brain weight. As before we write $Y$ for the vector of outcomes and $D$ for the matrix of covariates,
on which we condition throughout.
The model is given by
\begin{equation}
    \forall \: i \in \{1,\dotsc,n\}
    \quad
    y_i \mid \covariate_i,\beta,\sigma^2
    \sim
    \normal(\covariate_i\trans\beta , \sigma^2),\quad \prob(\beta,\sigma^2) \propto \sigma^{-2}
    \,,
\end{equation}
where $\sigma$ is a variance parameter, and $\beta\in\mathbb{R}^2$ is the regression coefficient. 
The training size is taken as $n_T = n/2 = 31$.
Exact posterior sampling on $\paramvar=(\beta,\sigma)$ is possible but we use Gibbs sampling instead for illustration purposes
following \cite{alqallaf2001cross}.
We initialize the chain by drawing $\beta_0, \beta_1 \sim \normal(0,1)$ independently and $\sigma^2 \sim \expo(1)$.
To obtain the full conditionals, we write the joint posterior density,
\begin{equation}
    \prob(\beta,\sigma^2 \mid Y)
    \propto
    (\sigma^2)^{-1-n/2}
    \exp\del{
        -\frac{1}{2\sigma^2}
        \norm{Y - \Covariate\beta}^2
    }
    \,.
\end{equation}
To get the conditional distribution of $\beta$ given $\sigma^2$ under the posterior distribution, note that
\begin{equation} \norm{Y-\Covariate\beta}^2 = (\beta - \hat{\beta})\trans (\Covariate\trans \Covariate) (\beta - \hat{\beta}) + \text{constant},\end{equation}
where $\hat{\beta} = (\Covariate\trans \Covariate)^{-1} \Covariate\trans Y$.
Thus, the conditional distribution is Normal with mean $\hat{\beta}$
and covariance matrix $\sigma^2 (\Covariate\trans \Covariate)^{-1}$.
The distribution of $\sigma^2$ given $\beta$ is inverse Gamma, where recall that
\begin{equation}
    \forall \: z\geq0 \quad\invgamma(z;a,b)
    = \frac{b^a}{\Gamma(a)} z^{-a-1}
      \exp\del{-\frac{b}{z}}
    \,.
\end{equation}
Then $\sigma^2$ given $\beta$ is inverse Gamma with $a = n/2$ and $b = \norm{Y-\Covariate\beta}^2/2$.
Coupling this algorithm can be done by maximal coupling of each of the conditional update of a Gibbs sampler.

In \citet{alqallaf2001cross}, the predictive performance in this example is measured by
the mean squared error, defined conditional on a split as
\begin{equation}
    r(\beta,\sigma^2)
    =
    \E
    \sbr{
        \norm{Y_V - Y_V^{\mathrm{pred}}}^2
        \mid
        Y_T, \beta, \sigma^2
    }
    \,,
\end{equation}
where $(T,V)$ denotes a data split, and $T = (D_T,Y_T)$, $V=(D_V,Y_V)$, recalling that there is an implicit conditioning on $D_T,D_V$ throughout this section.
Above, $Y_V^{\mathrm{pred}}$ is the predicted outcome,
and the expectation is taken with respect to the predictive distribution of $Y_V^{\mathrm{pred}}$
given $Y_T,\beta,\sigma^2$.
In this example this expectation is equal to $n_V \sigma^2
+ \norm{\Covariate_V \beta - Y_V}^2$.
Then one of the methods described in \citet{alqallaf2001cross} averages $r(\beta,\sigma^2)$
over MCMC draws approximating $\prob(\beta,\sigma^2 \mid Y_T)$.
Finally they average the results
across different random splits $(T,V)$.
We do not need unbiased path sampling to obtain an unbiased version of the above procedure:
we can readily use unbiased MCMC with the test function $h:(\beta,\sigma^2)\mapsto n_V \sigma^2 + \norm{\Covariate_V \beta - Y_V}^2$.
The proposed procedure reads: draw a partition $(T,V)$ randomly, and then obtain an unbiased estimator of $\pnorm(h)$ where $\pnorm$ is the posterior distribution given $\Covariate_T,Y_T$, and where $h$ is as above.

We implement this procedure and draw $1,000$ independent coupled chains.
We observe meeting times between $1$ and $5$.
Thus we set $k = 10$, $m = 25$, and draw $1,000$ independent unbiased estimators of $\text{CV}$.
We
obtain a $95\%$ confidence interval of $[32.79,33.03]$, and standard error of $0.06$.
By comparison, \citet{alqallaf2001cross} use 200 splits, and run 125 iterations of MCMC for each split, discarding the first 100.
The total number of Gibbs iterations performed is approximately the same, and \citet{alqallaf2001cross} obtain standard errors
that are similar.
An advantage of our method is in its simplicity: if we want more precise results, we simply
generate more independent estimators.

We now consider the criterion $-\log \prob(Y_V \mid Y_T)$, instead of the point-prediction mean squared error as above.
The sequence of distributions defined in \autoref{eq:cvsequence} is still amenable to a Gibbs sampling strategy and we
need to work out the conditional distributions.
The joint posterior density is
\begin{equation}
    \pnorm_\pathvar(\beta,\sigma^2)
    \propto
    (\sigma^2)^{-1-n_T/2 - \pathvar n_V/2}
    \exp\del{
        - \frac{1}{2\sigma^2}
        \norm{Y_T - \Covariate_T\beta}^2
        - \frac{\pathvar}{2\sigma^2}
        \norm{Y_V - \Covariate_V\beta}^2
    }
    \,.
\end{equation}
Note that
\begin{equation}
    \begin{split}
        &   -\frac{1}{2\sigma^2} \norm{Y_T - \Covariate_T\beta}^2-
        \frac{\pathvar}{2\sigma^2} \norm{Y_V - \Covariate_V\beta}^2\\
        &= -\frac{1}{2}(\beta - \mu_\pathvar)\trans \Lambda_\pathvar (\beta - \mu_\pathvar) + \text{constant},
    \end{split}
\end{equation}
with
\begin{equation}
    \begin{split}
    & \Lambda_\pathvar = \sigma^{-2} \Covariate_T\trans \Covariate_T + \pathvar \sigma^{-2} \Covariate_V\trans \Covariate_V  \\
    & \mu_\pathvar = \Lambda_\pathvar^{-1} \left(\sigma^{-2} \Covariate_T\trans Y_T + \pathvar \sigma^{-2} \Covariate_V\trans Y_V\right),
    \end{split}
\end{equation}
so that $\beta$ given the rest is $\normal(\mu_\pathvar, \Lambda_\pathvar^{-1})$.
On the other hand $\sigma^{2}$ given the rest is inverse Gamma $(a,b)$ with
\begin{equation}a = (n_T +\pathvar n_V)/2, \quad b = \norm{Y_T-\Covariate_T\beta}^2/2 + \pathvar \norm{Y_V - \Covariate_V \beta}^2/2.\end{equation}
This enables a Gibbs sampler targeting $\pnorm_\pathvar$ for any $\pathvar\in[0,1]$.
Next, we compute
$\nabla_\pathvar \log \unnorm_\pathvar$ as a function of $\beta,\sigma^2$, which we need to obtain UPS estimators.
We write
\begin{equation}\nabla_\pathvar \log \unnorm_\pathvar(\beta,\sigma^2) = \log \prob(Y_V \mid \Covariate_V,\beta,\sigma^2) =
\sum_{j=1}^{n_V} \log \varphi(y_{V,j}; \covariate_{V,j}\trans \beta, \sigma^2),\end{equation}
where $y_{V,j}$ refers to the $j$-th outcome in the validation set, $\covariate_{V,j}$ to the $j$-th row vector of corresponding covariates,
and $\varphi(z ; \mu,\sigma^2)$ is the Normal pdf evaluated at $z$, with mean $\mu$ and variance $\sigma^2$.
We observe that $\nabla_\pathvar \log \unnorm_\pathvar(\beta,\sigma^2)$ and powers of it are integrable
with respect to $\pnorm_\pathvar$, noting that $\pnorm_\pathvar$ is here a Normal-inverse-Gamma distribution.

This enables unbiased estimators of $-\log \prob(Y_V \mid Y_T)$, and thus of $\mathrm{CV}$ in \autoref{eq:cvobjective}.
Running $1,000$ independent estimators, we observe meeting times all less than $5$, and setting $k = 10$ and $m=25$,
we obtain an estimated CV criterion of $33.97$ with a standard error of $0.1$.

\subsubsection{Stack loss data \label{sec:stackloss}}

\begin{figure}
    \centering
    \begin{subfigure}[t]{0.6\textwidth}
     \includegraphics[width=\textwidth]{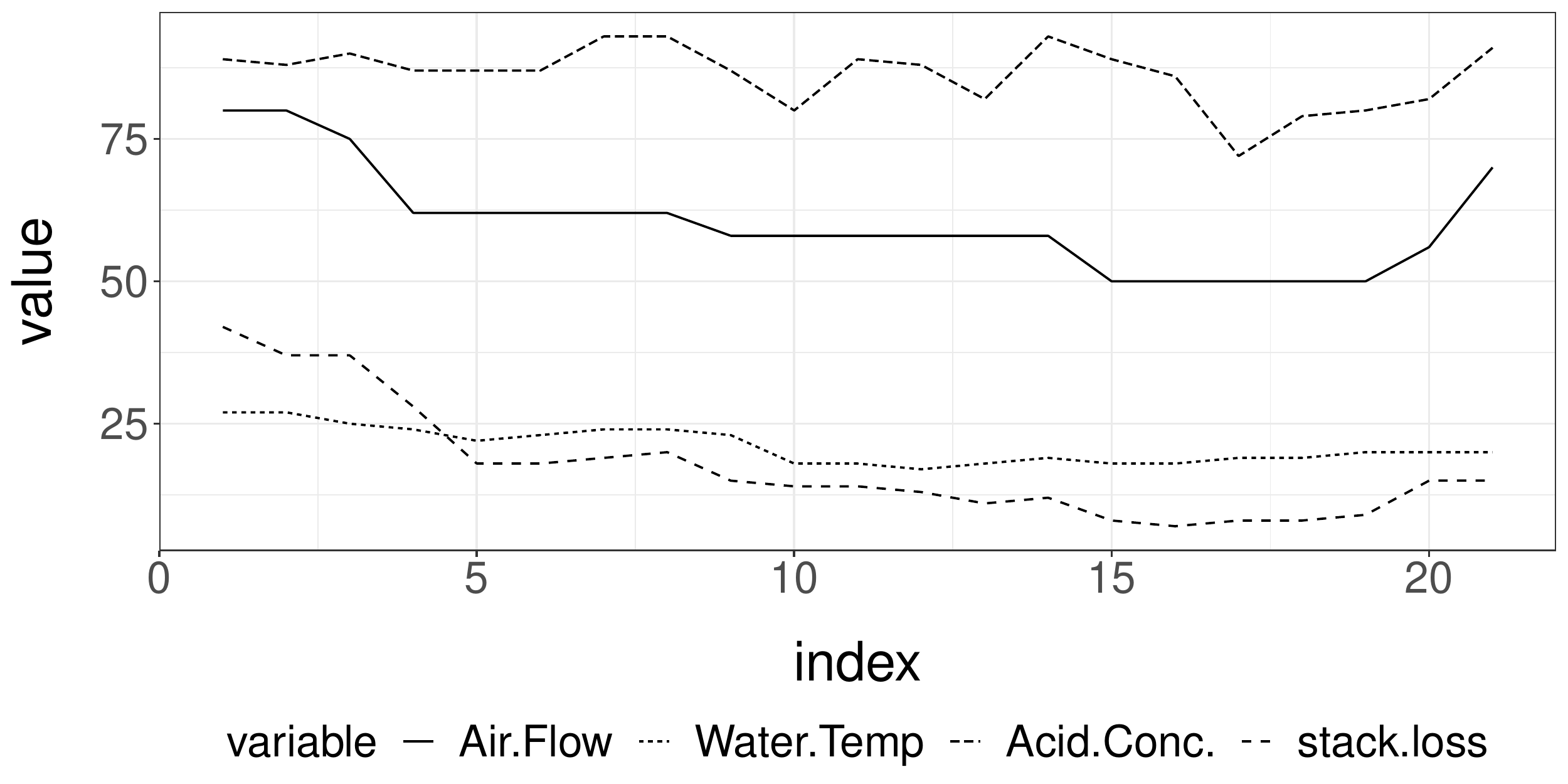}
    \caption{Stack loss data; \texttt{stack.loss}
    is the outcome, the other variables are covariates.}
    \label{fig:stackloss:data}
    \end{subfigure}
    \begin{subfigure}[t]{0.3\textwidth}
     \includegraphics[width=\textwidth]{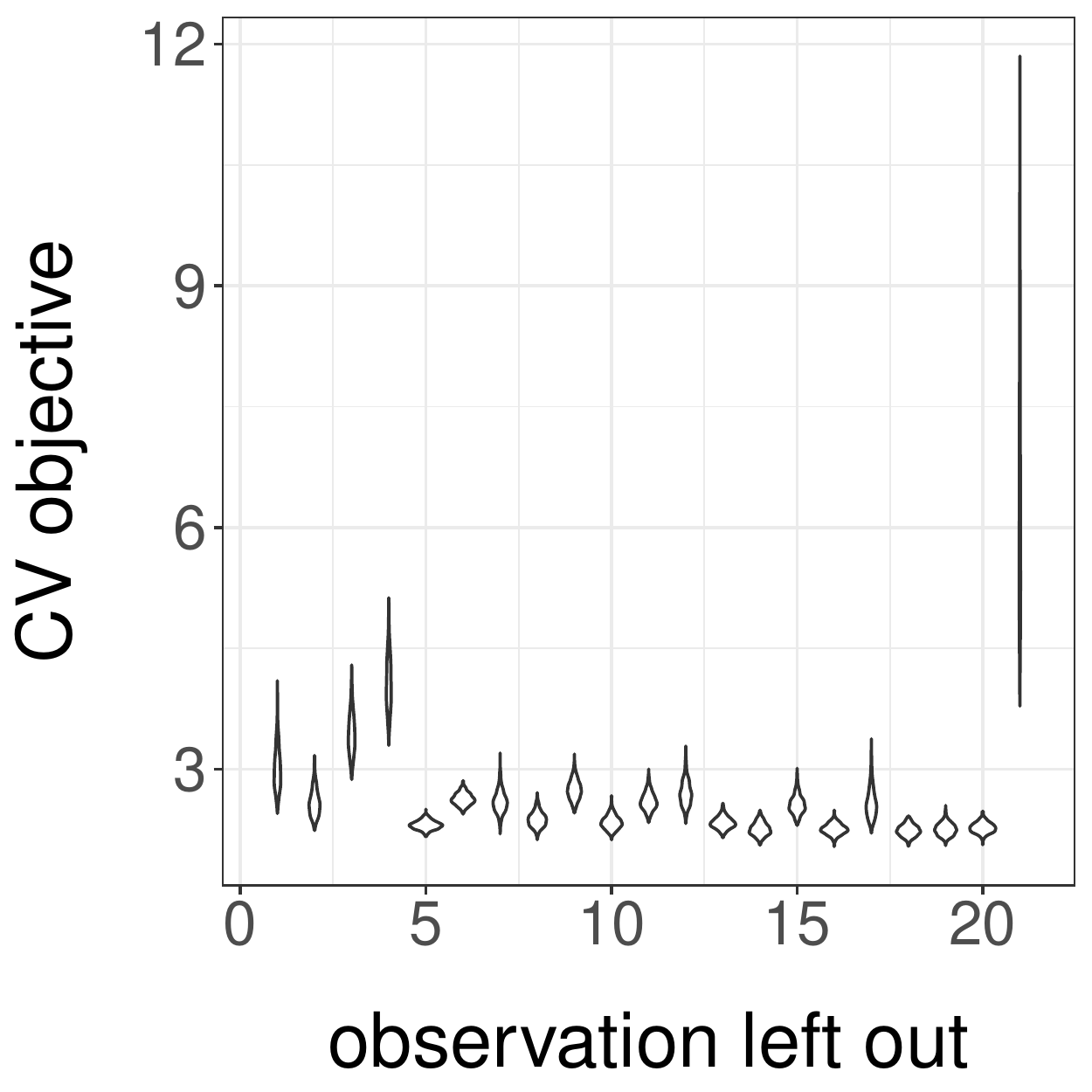}
    \caption{CV objective against index of left-out observation.}
    \label{fig:stackloss:cv}
    \end{subfigure}
    \caption{\label{fig:stackloss} Cross-validation (leave-one-out) for the linear regression on
    the stack loss data of \autoref{sec:stackloss}.
Data on the left, and CV objective against left-out observation on the right.
}
\end{figure}

We consider the stack loss data example, which was considered in \citet{peruggia1997variability,vehtari2017practical}.
In the former article, it is shown that importance sampling from the posterior given all the data to the posterior leaving one data point out
can lead to infinite variance estimators.
Here we use the \texttt{stackloss} data set of \citep{RCRAN},
with the outcome set to be the column \texttt{stack.loss}, and the covariates \texttt{Air.Flow}, \texttt{Water.Temp}, \texttt{Acid.Conc.}, and a column of ones.
The data are shown in \autoref{fig:stackloss:data}.
We consider leave-one-out cross-validation, with $n_T = n - 1 = 20$ here.
For simplicity we use the same model as in the previous section,
with a flat prior on $\beta$ given $\sigma^2$, instead of the proper prior given in \citet{peruggia1997variability}.

Using the coupled Gibbs sampler described in the previous section, we find meeting times
to be less than $10$ with large probability, thus we set $k = 10$ and $m=25$.
We obtain the CV estimators
shown in \autoref{fig:stackloss:cv}, based on $10,000$ independent replicates, plotted against the index of the left-out observation.
As in \autoref{sec:leukemia}, we can see that the variance of the CV estimators varies across the different ways of partitioning the data
into training and validation sets.
These CV estimators yield the $95\%$ confidence interval $[2.78, 2.82]$.

\section{Discussion \label{sec:discussion}}

Further work will be needed to compare the proposed estimators with state-of-the-art methods such as
sequential Monte Carlo samplers for normalizing constant estimation \citep[e.g.][]{lee2015variance,zhou2016toward,andrieu2016sampling},
with alternative approaches such as the ones described in \citet{chen1997monte,johnson1999posterior,neal2005estimating,salomone2018unbiased}
and references therein,
and with the different existing approaches for Bayesian cross-validation \citep[e.g.][]{alqallaf2001cross,bornn2010efficient,vehtari2017practical}.

Our estimators combine the path sampling identity with unbiased estimators of intractable integrals.
As such, they are expected to break if either path sampling or the unbiased estimators break.
Path sampling can give poor results if the path of distributions is ill-chosen, thus the design of these paths remains crucial.
We have seen in \autoref{sec:pgg} that different paths can give orders of magnitude differences in efficiencies.
We have also seen that the paths can benefit from approximations of the posterior distribution,
such as Laplace approximations. Mixtures of distributions
fitted on MCMC samples or variational approximations could also be considered. 
Conditional on a path, the choice of distribution $\probIS(\dif\pathvar)$ is also important and can be guided by preliminary runs.
Unbiased MCMC estimators themselves break either if the underlying MCMC algorithms mix poorly, or if the coupling strategy is ineffective;
we defer to \citet{jacob2017unbiased} for related discussions, and to \citet{heng2017unbiased} for the case of Hamiltonian Monte Carlo algorithms.

We note that the path sampling identity \autoref{eq:pathsamplingidentity} is an instance of a nested Monte Carlo (MC) problem, as defined and discussed in \citet{rainforth2016pitfalls}.
The target of nested MC is an expectation $I$ of the form
\begin{equation}
    \begin{split}
        & I = \Esup
            \sbr{
                f\del{
                    \pathvar,
                    \EU\del{\pathvar}
                }
            }\,\text{, with} \\
        & \EU\del{\pathvar} = \E_{\paramrv \sim \prob(\paramvar \mid \pathvar)}
            \sbr{
                \phi\del{\pathvar, \paramrv}
            }
        \,,
    \end{split}
    \label{eq:nestedMC}
\end{equation}
where the functions $f$, $\phi$ and the joint distribution of $(\paramvar,\pathvar)$ are problem-dependent choices.
In the case of path sampling, we obtain $I = \logratio$ by choosing:
\begin{equation}
\begin{split}
    & \prob(\pathvar) = \probIS(\pathvar) \,, \\
    & \prob(\paramvar \mid \pathvar) = \pnorm_\pathvar(\paramvar) \\
    & \phi\del{\pathvar, \paramvar} =
            \nabla_\pathvar
            \log\unnorm_\pathvar(\paramvar)
    \,\text{, and} \\
    & f(\pathvar, \EU\del{\pathvar}) = \EU\del{\pathvar} / \probIS(\pathvar)
    \,.
\end{split}
\end{equation}
In this case $f(\pathvar,\EU\del{\pathvar})$ is linear in its second argument, 
thus, given $\lambda$, unbiased estimators of $\EU\del{\pathvar}$ directly translate into unbiased estimators of $f(\pathvar,\EU\del{\pathvar})$.
We remark that unbiased estimators could also be obtained for functions $f$ that are nonlinear in the second argument.
For instance we can get an unbiased estimator of $\{\EU(\pathvar)\}^k$,
by sampling $k$ independent estimators of $\EUhat(\pathvar)$ and taking their product. More generally we can obtain unbiased estimators of $f(\pathvar,\EU\del{\pathvar})$ given $\pathvar$
for functions $f$ that are polynomials in the second argument. 

Finally it is possible to adapt the proposed approach to estimate the Bayesian cross-validation objective associated with some other scoring rules,
such as the one proposed in \citet{hyvarinen2005}, and considered in the setting of model comparison in e.g. \citet{dawid2015,shao2017bayesian}.

\subsubsection*{Acknowledgements}

The authors are grateful to Jeremy Heng and Stephane Shao for helpful discussions.

\bibliographystyle{apalike}
\bibliography{Biblio}

\end{document}